\documentclass[epj,fleqn]{svjour}

\usepackage[english]{babel}
\usepackage[T1]{fontenc} 
\usepackage{lmodern} 

\usepackage[pdftex]{graphicx,xcolor}
\usepackage{amsmath,amssymb}
\usepackage{mathrsfs,wasysym}
\usepackage{booktabs}
\usepackage{cancel}
\usepackage{cite}  
\usepackage[colorlinks = true,
            linkcolor = blue,
            urlcolor  = blue,
            citecolor = blue,
            anchorcolor = blue]{hyperref}
\numberwithin{equation}{section}
\usepackage[font=small,labelfont=bf,margin=0mm,labelsep=period,tableposition=top]{caption}
\newcommand{\kt}{k_t}
\newcommand{\kb}{k_t}
\newcommand{\MSbar}{\overline{\text{MS}}}

\newcommand{\as}{\alpha_s}
\newcommand{\asq}{\alpha_s(Q^2)}
\newcommand{\Ord}{\mathcal{O}}

\renewcommand{\Im}{\mathrm{Im}\:}
\let\originalleft\left
\let\originalright\right
\renewcommand{\left}{\mathopen{}\mathclose\bgroup\originalleft}
\renewcommand{\right}{\aftergroup\egroup\originalright}
\newcommand{\U}{\mathcal{U}}
\newcommand{\sqmatr}[4]{\left(
    \begin{array}[c]{cc}
      #1 & #2 \\ #3 & #4
    \end{array}
  \right)}

\def\beq{\begin{equation}}  
\def\eeq{\end{equation}}
\def\({\left(}
\def\){\right)}
\def\[{\left[}
\def\]{\right]}
\def\nn{\nonumber \\}
\allowdisplaybreaks

\newcommand{\new}[1]{{\textcolor{black}{ #1}}}

\abstract{Small-$x$ logarithmic enhancements arising from high-energy gluon emissions affect both the evolution
  of collinearly-factorized parton densities and partonic coefficient functions.
  With the higher collider energy reached by the LHC, the prospect of a future high-energy collider,
  and the recent deep-inelastic scattering (DIS) results at small-$x$ from HERA, providing phenomenological tools for
  performing small-$x$ resummation has become of great relevance.
  In this paper we discuss a framework to perform small-$x$
  resummation for both parton evolution and partonic coefficient
  functions and we describe its implementation in a computer code
  named High-Energy Large Logarithms (\texttt{HELL}). We present
  resummed and matched results for the DGLAP splitting functions and,
  as a proof of principle, for the massless structure functions in DIS.
  \new{Furthermore, we discuss the uncertainty from subleading terms on our results.}
  \\[2ex]
  OUTP-16-20P, Edinburgh 2016/13
}

\begin{document}

\title{ Small-$x$ resummation from HELL}

\author{Marco Bonvini,\inst{1}\thanks{\email{marco.bonvini@physics.ox.ac.uk}}
        \and
        Simone Marzani\inst{2}\thanks{\email{smarzani@buffalo.edu}}
        \and
        Tiziano Peraro\inst{3}\thanks{\email{tiziano.peraro@ed.ac.uk}}
}

\institute{Rudolf Peierls Centre for Theoretical Physics, University of Oxford, 1 Keble Road, Oxford, England, UK
          \and University at Buffalo, The State University of New York, Buffalo NY 14260-1500, USA
          \and Higgs Centre for Theoretical Physics, School of Physics and Astronomy, The University of Edinburgh, Edinburgh EH9 3JZ, Scotland, UK
}

\date{}
\maketitle

\tableofcontents

\section{Introduction}\label{sec:intro}

One aspect that makes the physics program of the CERN Large Hadron Collider (LHC) particularly rich is the vast kinematic region that can be explored. For inclusive enough processes, the kinematics is traditionally parametrized with a dimensionful scale $Q$, the typical hard scale of a process, e.g.\ a final-state invariant mass, and with the dimensionless ratio $x=Q^2/s$, with $\sqrt{s}$ the machine energy. 
Thus, the success of the LHC physics program relies upon having control of the many ingredients that enter theoretical predictions, over a wide kinematic range in both $x$ and $Q^2$. 
This includes high-order corrections in QCD and in the electro-weak sector, resummation effects and non-perturbative inputs to hadron-hadron cross section such as parton distribution functions (PDFs), which often represent the main source of theoretical uncertainty. 

The bulk of experimental data that constrain PDFs comes from deep-inelastic scattering (DIS) data collected by the HERA experiments~\cite{Abramowicz:2015mha}, which span several orders of magnitude in both $x$ and $Q^2$. 
Here, we concentrate on the high-energy, or small-$x$, regime.
In particular, at low $Q^2$, these data reach very small values of $x$, perhaps outside the region of validity of the fixed-order calculations which are used as inputs in the fits.
Moreover, in the context of LHC physics, the unique design of the LHCb detector (essentially a forward spectrometer) makes this experiment well-suited to access a region of phase-space of very large rapidities, thus providing useful data to pin down the largely unconstraint PDFs at small $x$. The success of this enterprise relies on having a reliable theory description of the low-$x$ region.

As we approach the small-$x$ regime, logarithms of $x$ become large and need to be resummed. As a consequence, PDF fits that are purely based on fixed-order matrix elements, may become unreliable at low $x$. Indeed, recent studies reveal some tension between low-$x$ and low-$Q^2$ data and standard fixed-order DGLAP fits~\cite{Caola:2009iy,Caola:2010cy,Abramowicz:2015mha}. 
High-energy logarithms appear both in partonic cross sections and in the DGLAP splitting functions~\cite{Gribov:1972ri,Dokshitzer:1977sg,Altarelli:1977zs}, which govern the evolution of the parton densities. The resummation of these contributions is based on the BFKL equation~\cite{Lipatov:1976zz,Fadin:1975cb,Kuraev:1976ge,Kuraev:1977fs,Balitsky:1978ic,Fadin:1998py}. However, it turns out that the correct inclusion of leading-logarithmic (LL) and next-to-leading logarithmic (NLL) corrections is far from trivial. 
This problem received great attention in 1990s, by more than one group, see, for instance, Refs.~\cite{Salam:1998tj,Ciafaloni:1999yw,Ciafaloni:2003rd,Ciafaloni:2007gf}, Refs.~\cite{Ball:1995vc,Ball:1997vf,Altarelli:2001ji,Altarelli:2003hk,Altarelli:2005ni,Altarelli:2008aj,Rojo:2009us}, and Refs.~\cite{Thorne:1999sg,Thorne:1999rb,Thorne:2001nr,White:2006yh}, which resulted in resummed anomalous dimensions for PDF evolution (for recent work in the context of effective theories, see~\cite{Rothstein:2016bsq}).

Small-$x$ resummation of partonic cross sections is based on the so-called $k_t$-factorization theorem~\cite{Catani:1990xk,Catani:1990eg,Collins:1991ty,Catani:1993ww,Catani:1993rn,Catani:1994sq,Ball:2007ra,Caola:2010kv}, which has been used to compute the high-energy behaviour of perturbative cross section for several processes such as heavy quark production~\cite{Ball:2001pq}, DIS~\cite{Catani:1994sq}, Drell-Yan~\cite{Marzani:2008uh}, direct photon~\cite{Diana:2009xv,Diana:2010ef} and Higgs production~\cite{Hautmann:2002tu,Pasechnik:2006du,Marzani:2008az,Caola:2011wq}. The formalism has been subsequently extended to rapidity~\cite{Caola:2010kv} and transverse momentum distributions~\cite{Forte:2015gve}.

Despite the wealth of calculations listed above, very few
phenomenological studies that incorporate both fixed-order and
resummed calculations exist.
The reason for this is technical: small-$x$ resummation requires an
all-order class of subleading corrections in order to lead to stable results.
The purpose of this paper is to remedy this deficiency. We develop a
framework to perform small-$x$ resummed phenomenology. Our starting
point is the resummation of coefficient and splitting functions according to the formalism developed by Altarelli, Ball and
Forte (ABF)~\cite{Ball:1995vc,Ball:1997vf,Altarelli:2001ji,Altarelli:2003hk,Altarelli:2005ni,Altarelli:2008aj}.
However, as we will describe in the paper, we introduce a number of
improvements that make the procedure easier to extend to new
processes, as well as numerically more stable.
For the first time, we make resummed splitting and coefficient functions available in a public code named \texttt{HELL} (High-Energy Large Logarithms).

The structure of this paper is the following.
In Sect.~\ref{sec:anom-dim} we describe the ABF resummation of the
splitting functions and its \texttt{HELL} implementation, highlighting
and motivating several improvements. 
%
\new{We then perform a comparison to the ABF original results and also to the ones of Ref.~\cite{Ciafaloni:2007gf}.}
In Sect.~\ref{sec:coeff-func} we
introduce a method to perform the resummation of coefficient functions directly
in transverse momentum space, which is then implemented in \texttt{HELL}. We show
its equivalence to the ABF Mellin-space resummation, while discussing
the numerous advantages of the new method. As a proof of principle, we
present results for the partonic coefficient functions of the massless DIS structure functions $F_2$ and
$F_L$, \new{as well as their comparison to the results obtained by ABF in Ref.~\cite{Altarelli:2008aj}.}

Finally, we draw our conclusions in Sect.~\ref{sec:conclusions}
and we outline forthcoming phenomenological studies which include fits
of PDFs, as well as studies of small-$x$ effects in electro-weak boson production
 at the LHC and Future Circular Colliders (FCC). Technical details are collected in a number of appendices.

\section{Resummation of DGLAP evolution kernels}
\label{sec:anom-dim}

In this section we review the construction of resummed DGLAP evolution kernels
needed for resummed PDF evolution up to NLL. We follow the formalism developed in the ABF series of papers~\cite
{Ball:1995vc,Ball:1997vf,Altarelli:2001ji,Altarelli:2003hk,Altarelli:2005ni,Altarelli:2008aj}.
\new{We will also comment about other approaches, but leave a thorough analytic comparison to future work.}
Most of the section is devoted to introducing notation and describing how the theoretical
results can be practically implemented in the code \texttt{HELL}.
We will also present several improvements over the original implementation.

It is convenient to work in the space of the variable $N$ conjugate by Mellin transformation to the variable $x$,
\beq\label{eq:MellDef}
f_i(N,Q^2) = \int_0^1dx\, x^N\, f_i(x,Q^2),
\eeq
since all convolutions become ordinary products.
Here $f_i(x,Q^2)$ is a generic PDF, and we used a non-standard notation for the Mellin transform
in which the kernel is $x^N$ rather than $x^{N-1}$. This is useful when discussing small-$x$
because the small-$x$ singularities, of the form $(1/x)\ln^kx$, are mapped into poles
in $N=0$ (in the usual notation, the poles are in $N=1$):
\beq\label{eq:logMellin}
  \int_0^1dx\, x^N\, \as^n \frac{ \ln^{k-1}x}{x}= (-1)^{k+1} (k-1)! \, \frac{\as^n}{N^{k}}.
\eeq
LL contributions at small-$x$ correspond to terms in Eq.~\eqref{eq:logMellin} with $k=n$ to all orders $n$ in $\as$,
while NLL ones have $k=n-1$. Note that double logarithmic corrections, which would correspond to $k=2n$, are absent in QCD,
with the noticeable exception of the Higgs production in gluon fusion with a pointlike effective vertex in the large-$m_t$ effective theory~\cite{Hautmann:2002tu}.

The dominant small-$x$ logarithmic enhancement only affects the singlet sector, while (double) logarithmic terms in the non-singlet are power-suppressed, i.e. they correspond to poles in $N=-1$. Therefore, we focus on the $2\times2$ singlet evolution matrix.
The construction of resummed anomalous dimensions, which are the Mellin transform of the splitting functions, can be divided into three successive steps:
\begin{enumerate}
\item resummation of the ``largest'' eigenvalue $\gamma_+$ of the singlet anomalous dimension matrix
\item resummation of the quark-sector anomalous dimension $\gamma_{qg}$
\item construction of the resummed anomalous dimension matrix in the physical (flavor) basis.
\end{enumerate}
We address these three steps in turn, giving a brief summary of the ABF procedure, emphasizing those aspects that are different from the original construction.
We finally comment on the numerical implementation and present some results.

\subsection{Resummation of the largest eigenvalue}
\label{sec:resgamma+}

The singlet-sector DGLAP evolution equation reads
\beq\label{eq:DGLAPmatrix}
Q^2 \frac{d}{d Q^2}\,
\begin{pmatrix}
f_g\\
f_q
\end{pmatrix}
= \Gamma\(N,\asq\)\,
\begin{pmatrix}
f_g\\
f_q
\end{pmatrix},
\eeq
where $f_g=f_g(N,Q^2)$ and $f_q=f_q(N,Q^2)$ are the gluon and quark-singlet PDFs respectively, and the
evolution matrix is given by (omitting arguments for readability)
\begin{equation}\label{eq:GammaDef}
  \Gamma(N,\as) \equiv
  \begin{pmatrix}
    \gamma_{gg} & \gamma_{gq} \\
    \gamma_{qg} & \gamma_{qq}
  \end{pmatrix}.
\end{equation}
As already mentioned, the non-singlet sector is not affected by small-$x$ logarithmic enhancement, and we therefore ignore it.

The DGLAP evolution equation Eq.~\eqref{eq:DGLAPmatrix} can be diagonalised by performing
a change of basis. We define the ``eigenvectors'' $f_\pm$ as
\beq\label{eq:f+-def}
\begin{pmatrix}
f_+\\
f_-
\end{pmatrix}
=
R\(N,\asq\) \,
\begin{pmatrix}
f_g\\
f_q
\end{pmatrix},
\eeq
where the transformation matrix $R$ (and its inverse) can be generically written as
\beq\label{eq:Rdef}
R =\frac1{r_--r_+}
\begin{pmatrix}
r_- & -1 \\
-r_+ & 1
\end{pmatrix},
\qquad
R^{-1} =
\begin{pmatrix}
1 & 1 \\
r_+ & r_-
\end{pmatrix}.
\eeq
Substituting Eq.~\eqref{eq:f+-def} into Eq.~\eqref{eq:DGLAPmatrix} we get
\beq\label{eq:DGLAPdiag}
Q^2 \frac{d}{d Q^2}\,
\begin{pmatrix}
f_+\\
f_-
\end{pmatrix}
= \[R\Gamma R^{-1} + Q^2 \frac{d R}{d Q^2} R^{-1} \]
\begin{pmatrix}
f_+\\
f_-
\end{pmatrix}.
\eeq
In general, to make the equation diagonal, one has to provide a matrix $R$ such that the matrix in squared brackets
in Eq.~\eqref{eq:DGLAPdiag} is diagonal,
\beq
R\Gamma R^{-1} + Q^2 \frac{dR}{d Q^2} R^{-1} =
\begin{pmatrix}
  \gamma_+ & 0 \\
  0 & \gamma_-
\end{pmatrix}.
\eeq
Solving this problem in general is rather complicated.
However, we notice that at pure LL level the matrix that diagonalizes $\Gamma$ has constant coefficients,
so we can ignore the second term in squared brackets and simply solve an eigenvalue problem.
At NLL, a non-trivial dependence on $Q^2$ appears; however, the action of the derivative with respect to
$Q^2$ further suppresses the second term in squared brackets by $\as\beta_0$, showing that it first contributes at NNLL level.
Therefore, when treating running coupling effects perturbatively, we can ignore the derivative contribution
and simply focus on the eigenvalue problem, which in particular leads to the following explicit form for $R$,
\beq\label{eq:r+-defFC}
r_\pm = \frac{\gamma_{qg}}{\gamma_\pm-\gamma_{qq}},
\eeq
being $\gamma_\pm$ the eigenvalues of $\Gamma$.
We anticipate that running coupling effects will eventually be resummed to all orders in $\as\beta_0$:
when this counting is adopted, the derivative term is no longer subleading and the matrix $R$ should be corrected for it.
We will come back to this point later in Sect.~\ref{sec:resmatrix} and in Sect.~\ref{sec:rotation}.

The eigenvalue $\gamma_+$ is chosen to be the largest eigenvalue at small-$x$, i.e.\ $N\sim0$,
namely the one which is enhanced at small $N$, while $\gamma_-$ is finite in $N=0$.
Consequently, $f_+$ is the only eigenvector that contains logarithmic enhancement and
which is affected by high-energy resummation.
This holds for several factorization schemes, including DIS and $\MSbar$, and the so-called $Q_0\MSbar$ scheme
which is particularly useful in small-$x$ resummation~\cite{Catani:1993ww,Catani:1994sq,Ciafaloni:2005cg,Marzani:2007gk}.
The resummation of small-$x$ logarithms in the evolution is then encoded in the
resummation of the largest eigenvalue $\gamma_+$.
\new{
The difference between the $\MSbar$ and $Q_0\MSbar$ factorization schemes influences the resummation of $\gamma_+$ beyond the leading logarithmic accuracy, as well as the resummation of $\gamma_{qg}$ and of the coefficient functions, as we shall see in more detail in Sec.~\ref{sec:coeff-func}. 
The structure of the resummation described in the remainder of the section is rather general and it is valid for both  $\MSbar$ and $Q_0\MSbar$ schemes. When presenting phenomenological results our scheme of choice will be $Q_0\MSbar$, which is preferred from an all-order viewpoint, because it gives more stable results~\cite{Altarelli:2008aj}.
It has to be noted that, when expanded to fixed-order, the difference between the two schemes only starts at relative $ \Ord(\as^3)$:
thus, all theoretical predictions that enter current PDF fits are not sensitive to this choice.}

High-energy resummation is achieved thanks to the BFKL equation~\cite
{Lipatov:1976zz,Fadin:1975cb,Kuraev:1976ge,Kuraev:1977fs,Balitsky:1978ic,Fadin:1998py}, which, in analogy with DGLAP, we
write as an evolution equation for the moments of the parton density. 
Therefore, defining the $M$ moments of $f_+$ by
\begin{equation}
  f_+(x,M) = \int_{-\infty}^\infty\, \frac{d Q^2}{Q^2} \; \(\frac{Q^2}{Q_0^2}\)^{-M}\, f_+(x,Q^2),
\end{equation}
with $Q_0$ some reference scale (the PDFs depend logarithmically on $Q$, so the value of $Q_0$ is irrelevant), we have
\begin{equation}\label{eq:bfkl}
 -x\, \dfrac{d}{dx}\, f_+(x,M)=\chi(M,\as)\, f_+(x,M),
\end{equation}
where $\chi$ is the BFKL kernel, currently known to NLO~\cite{Fadin:1998py} and to NNLO in the collinear approximation~\cite{Marzani:2007gk} (see Ref.~\cite{Caron-Huot:2016tzz} for recent work beyond NLO accuracy).
\new{In the small-$x$ and high-$Q^2$ limit, both the DGLAP and BFKL equations are expected to hold,
and consistency between the solutions to both equations allows to resum to all orders
collinear contributions in the BFKL kernerl or, equivalently, small-$x$ contributions in the DGLAP anomalous dimension.}
Knowledge of the BFKL kernel to N$^k$LO accuracy allows for the resummation of the N$^k$LL contributions to the DGLAP anomalous dimension \new{(and vice-versa)}.
It is worth noting that Eq.~\eqref{eq:bfkl} is an ordinary differential equation only if the coupling does not run.
Indeed, in $M$-space, $\as(Q^2)$ becomes a differential operator $\hat{\alpha}_s$, essentially because 
$\ln Q^2$ is turned into $-\partial/\partial M$ and consequently Eq.~\eqref{eq:bfkl} is to be intended as an operator-valued equation.
This is a manifestation of the well-known fact that the eigenvalues of the LO kernel do not diagonalize the BFKL equation at NLO. 

\new{Consistency between DGLAP and BFKL equations} allows us to build a \emph{double-leading} (DL)
expansion of $\gamma_+$ and $\chi$ which takes into account the
logarithmically enhanced contributions in both $\ln Q^2$ and
$\ln (1/x)$~\cite{Salam:1998tj}.  Because of the poor perturbative behaviour of the BFKL
kernel, obtaining a stable resummed result is however not
straightforward and requires a somehow complex procedure with a
careful treatment of the formally subleading terms.  
\new{This issue received great attention in the past, mainly by three groups: Refs.~\cite{Salam:1998tj,Ciafaloni:1999yw,Ciafaloni:2003rd,Ciafaloni:2007gf}, Refs.~\cite{Ball:1995vc,Ball:1997vf,Altarelli:2001ji,Altarelli:2003hk,Altarelli:2005ni,Altarelli:2008aj,Rojo:2009us}, and Refs.~\cite{Thorne:1999sg,Thorne:1999rb,Thorne:2001nr,White:2006yh}. 
Despite the different approaches, which are characterized by different treatments of formally subleading corrections, a fairly consistent picture emerged, with small differences in the final results of the different groups (see e.g.\ Ref.~\cite{Dittmar:2009ii}).}

The ABF approach~\cite{Ball:1995vc,Ball:1997vf,Altarelli:2001ji,Altarelli:2003hk,Altarelli:2005ni,Altarelli:2008aj},
which we adopt in this paper with a few improvements, allows us to build perturbatively stable resummed results by combining four main
ingredients: duality, symmetrization, 
momentum conservation and running coupling resummation, as we summarize below.

\emph{Duality} between the DGLAP anomalous dimensions and the BFKL
evolution kernel, is the statement that in the fixed coupling limit (i.e.\ neglecting
contributions due to the running of $\as$), the kernels satisfy the following relations~\cite{Jaroszewicz:1982gr,Catani:1989sg}
\begin{equation} \label{eq:duality}
  \chi(\gamma_+(N,\as),\as) = N  \quad\leftrightarrow\quad
\gamma_+(\chi(M,\as),\as) = M.
\end{equation}
Beyond LL Eq.~\eqref{eq:duality} is corrected by
contributions due to the running of $\as$.
In principle Eq.~(\ref{eq:duality}) provides all the ingredients for small-$x$ resummation: we start with the BFKL kernel $\chi$ at a given order (LO or NLO) and we use duality to determine a DGLAP anomalous dimension, dual to $\chi$, which resums small-$x$ contributions to the desired logarithmic accuracy (LL or NLL). 
However, as previously mentioned, the BFKL kernel itself
exhibits a very poor perturbative behaviour, with poles of the form
$\as^k/(j-M)^k$ for any integer $j$ at every perturbative order
$k$.  The poles in $M=0$ and $M=1$, which correspond to the collinear and anti-collinear regions, are particularly harmful~\cite{Salam:1998tj}. 
The key observation is that the resummation of collinear poles (which in momentum space are just collinear logarithms) is controlled by the DGLAP anomalous dimension. Hence, we can use duality, in the opposite direction, to derive a kernel $\chi$, dual to standard DGLAP, that resums all the collinear enhancements. 
The DL kernel can then be constructed by matching standard BFKL with the collinearly improved one. Furthermore, again by duality, this result can be turned into an anomalous dimension.

However,  the stabilization of the collinear region does not completely cure the problem,
because of the singularity of the BFKL kernel in $M=1$.
Indeed the behaviour in middle region between $M=0$ and $M=1$ determines \new{by duality} the nature of the rightmost
small-$N$ singularity, i.e.\ the asymptotic
small-$x$ behaviour of the splitting functions.  The nature of the singularity obtained in this way is perturbatively
unstable: it is a pole at fixed
order, a square root branch-cut at DL-LO, non-singular at DL-NLO, see e.g.~\cite{Altarelli:1999vw}.
The anticollinear terms can however be resummed and thus stabilized by
exploiting the symmetry properties of the BFKL kernel, which relate them to
the collinear contributions~\cite{Salam:1998tj, Altarelli:2005ni}.  This \emph{symmetrization} is performed
by constructing a kernel which coincides with the DL one at a
given logarithmic accuracy in $\ln Q^2$ and $\ln (1/x)$, but satisfies
the required symmetry properties exactly (while in general these would
be spoiled by subleading terms).  In the ABF approach, the symmetrized
kernel is defined via implicit equations which must be solved
numerically (more details are given in App.~\ref{app:gammaplusresdetails}).
\new{Note that the definition of the symmetrized kernels has some degree of arbitrariness, due to the inclusion of unconstrained subleading contributions.}
After symmetrization, the singular behaviour of the dual DGLAP anomalous dimension
is always a square root branch-cut.

The third  important ingredient of the ABF resummation
is \emph{momentum conservation}, which implies that the first Mellin moment of the largest eigenvalue must vanish,
and translates by duality into a constraint on the BFKL kernel:
\begin{equation} \label{eq:momcons}
  \gamma_+(1,\as) = 0 \qquad\to\qquad \chi(0,\as) = 1.
\end{equation}
In general, in a DL expansion, Eq.~\eqref{eq:momcons} is violated by
subleading terms, but it may be enforced by adding a subleading
contribution which does not introduce new singularities at small $N$
and vanishes at large $N$.
\new{We stress that, while the final anomalous dimension must clearly satisfy momentum conservation, one can decide whether momentum conservation should be
imposed in any of the intermediate steps. We note that}
the stability of the result is greatly
improved by enforcing momentum conservation at each step of the
resummation procedure.

Symmetrization and momentum conservation allow us to build perturbatively
stable BFKL kernels, and, by duality, DGLAP anomalous dimensions, in the fixed
coupling limit.  The resulting singularity is however modified at every perturbative order by
running coupling corrections to duality~\cite{Altarelli:2001ji}.  
These corrections start at NLL and,
while formally subleading, they are in fact dominant since they change
the nature of the small-$N$ singularity.  The dominant running
coupling corrections are resummed by solving the running-coupling BFKL evolution equation
for $f_+$, \new{and extracting its anomalous dimension (see e.g. Refs~\cite{Thorne:1999sg,Thorne:1999rb,Thorne:2001nr}).}
This can be done analytically by approximating the kernel
in proximity of its minimum, which in turn corresponds by duality to the
square-root branch cut of the anomalous dimension, i.e.\ its leading
singularity.  After \emph{running coupling resummation}, the rightmost
singularity of the anomalous dimension is turned back to a simple pole (as it was at fixed leading order), but now
shifted from $N=0$ to $N=N_B(\as)>0$.  The overall effect is a
suppression of the small-$x$ growth with respect to the (symmetrized)
DL result.

Combing all the effects together, the final form of the resummed DGLAP eigenvalue in the ABF approach at LO+LL is
\begin{align} \label{eq:gammaLOrc}
\gamma^{\text{LO}+\text{LL}}_{+}(N,\as)&=
   \gamma_+^{\Sigma,\text{LO}}(N,\as)+\gamma^{B,\rm LL}(N,\as) \nonumber \\&\quad-\gamma^\text{LO,LL d.c.}
  \end{align}
while at NLO+NLL is
\begin{align}\label{eq:gammaNLOrc}
  \gamma^{\text{NLO}+\text{NLL}}_{+}(N,\as) &=
   \gamma^{\Sigma,\text{NLO}}_{+}(N,\as)+\gamma^{B,\rm NLL} (N,\as)\nonumber \\&\quad
   -\gamma^\text{NLO,NLL d.c.}
\end{align}
In the above equations $\gamma_+^{\Sigma,\text{(N)LO}}$ contains the
symmetrized double-leading contributions at LO and NLO respectively, which include the fixed-order part of the anomalous dimensions.
The ``Bateman'' contribution $\gamma^{B,\rm (N)LL}$ contains the running coupling effects obtained by solving the evolution equation,
and carries the actual small-$N$ singularity.
The remaining term in each equation avoids double counting.
Further details and explicit formulas are given in App.~\ref{app:gammaplusresdetails}.
For later convenience, we also define 
\begin{align}\label{delta_gammma_+}
\Delta \gamma_+^\text{LL} &=   \gamma^{\text{LO}+\text{LL}}_{+}(N,\as)   -\gamma^{\text{LO}}_{+}(N,\as), \nonumber \\
\Delta \gamma_+^\text{NLL} &=   \gamma^{\text{NLO}+\text{NLL}}_{+}(N,\as)   -\gamma^{\text{NLO}}_{+}(N,\as),
\end{align}
which contain only the resummed contributions to be added to the corresponding fixed order. Note that one could also imagine to match the resummation to NNLO. This step, which is usually straightforward, is rather cumbersome in this case essentially because the dependence on the strong coupling of symmetrized DL result $\gamma_+^\Sigma$ is only known numerically. 
We leave this further matching for future work, stressing that it is of great interest especially in the context of PDF fits.

\new{Before moving to the resummation of the quark and gluon entries of the anomalous dimension matrix, let us briefly comment about the different approaches to the resummation of $ \gamma_+$ that can be found in the literature. 
The resummation proposed in Refs.~\cite{Salam:1998tj,Ciafaloni:1999yw,Ciafaloni:2003rd,Ciafaloni:2007gf} is based on very similar ingredients as ABF, namely the resummation of collinear singularities, symmetrization and running coupling effect. However, rather than relying upon duality to determine the resummed anomalous dimension, the running coupling BFKL equation is solved and the anomalous dimension is extracted from the solution.
Refs.~\cite{Thorne:1999sg,Thorne:1999rb,Thorne:2001nr,White:2006yh} on the other hand, only relied on running coupling corrections and not on symmetrization, with the argument that high-$Q^2$ physics should be dominated by the $M\sim0$ region.
A thorough study of all the sources of uncertainty in small-$x$ resummation of $\gamma_+$ would require investigating all the subleading modifications described above and goes beyond the scope of this work. However, given the conclusions of Ref.~\cite{Dittmar:2009ii}, which found the three different approaches to be in reasonable agreement, one might expect the small-$x$ resummation of the eigenvalue $\gamma_+$ to be under good control.}

\subsection{Resummation of the quark anomalous dimension}
\label{sec:resgammaqg}

The high-energy behaviour of the $qg$ anomalous dimension has been
derived at the leading logarithmic level in Refs.~\cite{Catani:1993rn, Catani:1994sq}.
The quark anomalous dimensions are always suppressed by a power of $\as$ with respect
to the gluon ones, so they enter for the first time at NLL.

The all-order small-$N$ behaviour of $\gamma_{qg}$ is determined from the resummed anomalous dimension $\gamma_+$:
\beq\label{eq:gamma_qg_h}
\gamma_{qg}(N,\as) = \as h\(\gamma_+(N,\as)\).
\eeq
In order to perform the resummation, the function $h$ to all orders in its argument is needed; however,
to the best of our knowledge, a closed form for $h$ in either $\MSbar$ or $Q_0 \MSbar$ does not exist.
Nevertheless, the coefficients $h_k$ of its Taylor expansion
\beq\label{eq:hqgser}
h(M) = \sum_{k=0}^\infty h_k M^k
\eeq
can be computed recursively, as described in Ref.~\cite{Catani:1994sq}.
The first $35$ coefficients have been worked out in Ref.~\cite{Altarelli:2008aj}.
The singular behaviour of $\gamma_{qg}$ up to $\Ord(\as^k)$ is obtained by including
the singular behaviour of $\gamma_+$ up to at least the same order.

We first address the question of which accuracy is needed for $\gamma_+$ in Eq.~\eqref{eq:gamma_qg_h}. 
Since NLL effects in $\gamma_+$ will contribute to NNLL in $\gamma_{qg}$, we could
use the LL expression for the largest eigenvalue.
However, since the position of the pole determines the asymptotic small-$x$ behaviour of the result,
the use of the LL $\gamma_+$ pole is not ideal because it would lead to
displaced poles in different entries of the anomalous dimension matrix.
Therefore, we find it convenient (mostly from a numerical point of view) to use an hybrid expression
which we denote LL$^\prime$ which is based on the DL-LO result but contains the running-coupling NLL contribution.
In formulae, we define
\begin{align}\label{eq:gammaLOp}
  \gamma_+^{\text{LO+LL}'}&= \gamma_+^{\Sigma,\text{LO}}(N,\as)+\gamma^{B,\rm NLL}(N,\as)\nonumber \\ &\quad -\gamma^\text{LO,NLL d.c.}.
\end{align}
In other words, this expression is basically the same as $\gamma_+^{\text{LO+LL}}$, Eq.~\eqref{eq:gammaLOrc},
but the parameters entering the Bateman
anomalous dimension $\gamma^B$ (and consequently all the double counting terms),
which determine the position of the pole, are those of the NLL result Eq.~\eqref{eq:gammaNLOrc}.

The function $\gamma_+^{\text{LO+LL}'}$, Eq.~\eqref{eq:gammaLOp}, cannot be directly used in
Eq.~\eqref{eq:gamma_qg_h}, because its growth at large $N$ (due to its fixed-order component)
would produce a spurious large $N$ behaviour in $\gamma_{qg}$ to all orders in $\as$.
Therefore, we use
\begin{equation}\label{eq:gammaLLp}
\gamma_+^{\text{LL}'}= \gamma_+^{\text{LO}+\text{LL}'}- \gamma_+^{\rm LO} + \gamma_+^{\rm LO,sing},
\end{equation}
where $\gamma_+^{\rm LO,sing}$ is the singular $N\sim0$ part of the LO anomalous dimension.%
\footnote{In principle it would be sufficient to include in only the singular LL contributions.
However, one might argue that it is safer to also include additional subleading (NLL) terms
in it, provided they vanish at large $N$.
We indeed include these NLL terms; details are given in App.~\ref{app:gammaqgcomparison}.}
We point out that this procedure differs from that of Ref.~\cite{Altarelli:2008aj}, where $h$ is computed
with $\gamma_+^{\rm NLO+NLL}$, and the large-$N$ behaviour is subtracted by recomputing
$h$ with $\gamma_+^{\rm NLO}-\gamma_+^{\rm NLO,sing}$.
We comment on the differences between the two approaches in App.~\ref{app:gammaqgcomparison}.
Here we just stress that the two procedures are formally equivalent,
our formulation leading to a faster and more reliable numerical implementation.

The resummation of running coupling contributions also affects the determination of $\gamma_{qg}$.
In the approach of Ref.~\cite{Ball:2007ra}, it is included by computing
\new{
\beq\label{eq:gamma_qg_ss_rc}
\gamma^{\rm NLL}_{qg} = \as \sum_{k=0}^\infty h_k \[\(\gamma_+^{\rm LL'}\)^k\],
\eeq
where the square brackets notation
$\[\gamma^k\]$ is defined by the recursion}
\beq\label{eq:recursion}
\[\gamma^{k+1}\] = \gamma \(1+k\frac{\dot\gamma}{\gamma^2}\)\[\gamma^k\], \qquad \[\gamma\] = \gamma,
\eeq
the dot denoting the derivative with respect to $\ln Q^2$. In our implementation, $\dot\gamma$ is computed as a derivative with respect to $\as$,
$\dot\gamma = -\beta_0\as^2 \partial\gamma/\partial\as$.
The need to compute a derivative with respect to $\as$ of the resummed anomalous dimension
is one of the main practical motivation for using $\gamma^{\rm LL'}_+$ rather than $\gamma^{\rm NLL}_+$,
as the numerical evaluation of the former is much faster and more stable than the latter,
thereby allowing a more precise determination of the numerical derivative.
\new{
Note that Eq.~\eqref{eq:recursion} comes from the approximate assumption
that $\gamma$ is linear in $\as$~\cite{Ball:2007ra} (we will explicitly re-derive this result in the context of coefficient functions in  in Sect.~\ref{sec:ABFcomparison}).  Under this assumption, $\dot\gamma$ would simply be $\dot\gamma \simeq -\beta_0\as \gamma$.
We shall also consider this additional approximate form for $\dot\gamma$ as a means to estimate the uncertainty due to this approximation.}

A further complication arises from the fact that after the inclusion of running coupling corrections Eq.~\eqref{eq:gamma_qg_ss_rc},
the series Eq.~\eqref{eq:hqgser} is only asymptotic.
In Ref.~\cite{Altarelli:2008aj} the resummation is performed by computing the sum of the series \`a la Borel,
using a truncated Borel integral corrected with an asymptotic behaviour derived from a simpler solvable model.
We adopt here a different approximate procedure, which only relies on the available information from $h$.
We make use of a Borel-Pad\'e summation procedure,
where we compute the sum of the series \`a la Borel, and use a Pad\'e approximant
for the sum of the Borel-transformed series obtained from a finite number of coefficients
of the expansion of $h$.
Details of this procedure are given in App.~\ref{app:Borel}.

Finally, from Eq.~\eqref{eq:gamma_qg_ss_rc} we can construct the pure resummed contribution
\beq\label{eq:deltagammqg}
\Delta\gamma_{qg}^{\rm NLL} = \gamma^{\rm NLL}_{qg} - \as h_0 - \as^2 h_1 \gamma_+^{\rm LO,sing}
\eeq
as the contribution to be added to the NLO anomalous dimension to obtain a matched NLO+NLL result.

\subsection{Construction of the resummed singlet splitting function matrix}
\label{sec:resmatrix}

Now that we have resummed the largest eigenvalue and the $qg$ component, we can construct the full
anomalous dimension matrix in the gluon-singlet basis.
First of all, the $qq$ component can be recovered by making use of the color-charge relation \cite{Catani:1994sq}
\beq\label{eq:gamma_qq}
\gamma_{qq}^{\rm NLL} = \frac{C_F}{C_A} \[\gamma_{qg}^{\rm NLL} - \frac{\as}{\pi}\frac{n_f}{3}\].
\eeq
The eigenvalue $\gamma_-$, which is finite in $N=0$ and does not resum,
contains a finite fixed-order constant terms which is formally NLL,
\beq\label{eq:gamma_-}
\gamma_-^{\rm NLL} = -\frac{\as}{\pi}\frac{n_f}{3} \frac{C_F}{C_A} .
\eeq
This particular form, together with the color-charge relation Eq.~\eqref{eq:gamma_qq},
is such that the $r_-$ component of the transformation matrix, Eq.~\eqref{eq:r+-defFC},
is simply given at NLL by $r_- = -C_A/C_F$, and is therefore $Q^2$-independent.

The $gg$ component can be recovered by transforming back the diagonal matrix to the physical basis,
leading to the general expression 
\beq
\gamma_{gg} = \frac{\gamma_+ - \gamma_-r_+/r_-}{1-r_+/r_-}.
\eeq
Using Eq.~\eqref{eq:r+-defFC}, valid in the fixed-coupling case, we simply get
$\gamma_{gg} = \gamma_++\gamma_--\gamma_{qq}$, which combined with Eq.~\eqref{eq:gamma_qq} and Eq.~\eqref{eq:gamma_-} leads to
\beq\label{eq:gamma_gg}
\gamma_{gg}^{\rm NLL} = \gamma_+^{\rm NLL} - \frac{C_F}{C_A} \gamma_{qg}^{\rm NLL}.
\eeq
When resumming running coupling effects, the form of $r_\pm$ changes
and, consequently, Eq.~\eqref{eq:gamma_gg} receives in principle
running-coupling corrections. However, we have checked that these
effects are typically smaller than the various sources of ambiguity
in the whole resummation procedure coming from subleading
contributions. Therefore, following Ref.~\cite{Altarelli:2008aj}, and
without loss of accuracy, we adopt Eq.~\eqref{eq:gamma_gg} as our
default implementation for $\gamma_{gg}$.
On the other hand, a more careful treatment of running
coupling effects is needed when dealing with the resummation of
coefficient functions, as we shall see
later in Sect.~\ref{sec:rotation}.

Finally, it remains to compute $\gamma_{gq}$; however, the available information is not sufficient to constraint its NLL part.
This is not a problem, because the accuracy of the solution of the evolution equation is formally NLL even if the $gq$
entry is just LL. At LL, we can just use a color-charge relation
\beq\label{eq:gamma_gq}
\gamma_{gq}^{\rm LL} = \frac{C_F}{C_A} \gamma_{gg}^{\rm LL};
\eeq
this equation can be modified by using the NLL expression of $\gamma_{gg}$, even though
the resulting $gq$ anomalous dimension will still remain formally accurate at LL.

For phenomenological application we find useful to write the resummed and matched anomalous dimensions as a fixed-order contribution
$\gamma^\text{(N)LO}$ plus a $\Delta \gamma^\text{(N)LL}$, which contains the resummation minus double counting.
Thus, in this notation, the NLO+NLL evolution matrix is given by
\beq\label{eq:NLLevolmatrix}
\Gamma^{\rm NLO+NLL} = 
\sqmatr
{\gamma_{gg}^{\rm NLO}}
{\gamma_{gq}^{\rm NLO}}
{\gamma_{qg}^{\rm NLO}}
{\gamma_{qq}^{\rm NLO}}
+ \sqmatr
{\Delta\gamma_{gg}^{\rm NLL}}
{\frac{C_F}{C_A}\Delta\gamma_{gg}^{\rm NLL}}
{\Delta\gamma_{qg}^{\rm NLL}}
{\frac{C_F}{C_A}\Delta\gamma_{qg}^{\rm NLL}},
\eeq
where $\Delta\gamma_{qg}^{\rm NLL}$ is given in Eq.~\eqref{eq:deltagammqg},
and $\Delta\gamma_{gg}^{\rm NLL}=\Delta\gamma_+^{\rm NLL}-(C_F/C_A) \Delta\gamma_{qg}^{\rm NLL}$ can be easily derived from Eq.~\eqref{eq:gamma_gg}.
From the above matrix, one can compute the inverse Mellin transform and obtain
the resummed splitting functions,
\beq\label{eq:xNLLevolmatrix}
P^{\rm NLO+NLL} = 
\sqmatr
{P_{gg}^{\rm NLO}}
{P_{gq}^{\rm NLO}}
{P_{qg}^{\rm NLO}}
{P_{qq}^{\rm NLO}}
+ \sqmatr
{\Delta P_{gg}^{\rm NLL}}
{\frac{C_F}{C_A}\Delta P_{gg}^{\rm NLL}}
{\Delta P_{qg}^{\rm NLL}}
{\frac{C_F}{C_A}\Delta P_{qg}^{\rm NLL}},
\eeq
where $\Delta P_{gg}^{\rm NLL}$ and $\Delta P_{qg}^{\rm NLL}$ are the ultimate primary ingredients for a resummed DGLAP evolution.

The results in momentum space deserve further comments. The contributions $\Delta \gamma_{ij}$ vanish, by construction, at large $N$. This is enough to guarantee that their $x$-space conjugates $\Delta P_{ij}$ are ordinary functions, i.e.\ they do not contain plus distribution or delta functions. However, they potentially exhibit a constant behaviour, or even an integrable singularity, as $x\to 1$.
To avoid potential problems with matching at large $x$, we follow ABF and we further suppress these functions at $x=1$
with an $x$-space damping:
\begin{subequations}\label{eq:damping}
\begin{align}
  \Delta P_{gg}^{\rm NLL} &\to (1-x)^2 \Delta P_{gg}^{\rm NLL} \\
  \Delta P_{qg}^{\rm NLL} &\to (1-x)^2 \Delta P_{qg}^{\rm NLL}.
\end{align}
\end{subequations}
However, despite the many desirable features of the above damping procedure,
momentum is no longer conserved in Eq.~\eqref{eq:NLLevolmatrix}.
In the flavor basis, momentum conservation implies that
\beq
\gamma_{gg}(1) + \gamma_{qg}(1) = 0,\qquad
\gamma_{gq}(1) + \gamma_{qq}(1) = 0,
\eeq
namely, the sum of each column must vanish in $N=1$.
Both equations imply
\begin{align}
&\Delta\gamma_{gg}^{\rm NLL}(1) + \Delta\gamma_{qg}^{\rm NLL}(1) = \nonumber\\
&\qquad \qquad \Delta\gamma_+^{\rm NLL}(1) + \(1-\frac{C_F}{C_A}\)\Delta\gamma_{qg}^{\rm NLL}(1) = 0,
\end{align}
which is violated.
The origin of the violation is twofold: first, even though $\Delta\gamma_+^{\rm NLL}$ is originally constructed to
vanish in $N=1$, it looses this property once the $x$-space damping is applied;
second, $\Delta\gamma_{qg}^{\rm NLL}$ is not necessarily vanishing at $N=1$ even in absence of damping.
While this momentum violation was not considered in the original ABF work~\cite{Altarelli:2008aj},
here we force momentum conservation by a modification of the $gg$ entry
\beq\label{eq:gamma_gg_momcons}
\Delta\gamma_{gg}^{\rm NLL} (N) = \Delta\gamma_+^{\rm NLL} (N) - \frac{C_F}{C_A}\Delta\gamma_{qg}^{\rm NLL} (N) - c\, d(N),
\eeq
where
\beq\label{eq:MCcoeff}
c = \frac{\Delta\gamma_+^{\rm NLL}(1) +\(1- \frac{C_F}{C_A}\)\Delta\gamma_{qg}^{\rm NLL}(1)}{d(1)},
\eeq
and $d(N)$ is any function that goes to zero at large $N$ and has no leading singularities.
We use
\beq
d(N) = \frac1{N+1}-\frac2{N+2}+\frac1{N+3} \; \leftrightarrow\; d(x) = (1-x)^2,
\eeq
so that the momentum conservation can be restored directly in both $N$ and $x$ space.

Given the numerous steps involved in the resummation procedure, we find useful to summarize our strategy in implementing them in a numerical code:
\begin{itemize}
\item we compute $\Delta\gamma_+^{\rm NLL}$ and $\Delta\gamma_{qg}^{\rm NLL}$ as described earlier in this Sects.~\ref{sec:resgamma+} and~\ref{sec:resgammaqg}, respectively;
\item we construct $\Delta\gamma_{gg}^{\rm NLL}$ from Eq.~\eqref{eq:gamma_gg};
\item we compute the inverse Mellin transforms $\Delta P_{gg}^{\rm NLL}$ and $\Delta P_{qg}^{\rm NLL}$;
\item we apply the damping Eqs.~\eqref{eq:damping};
\item we compute the $N=1$ Mellin moments of the damped functions, and construct $c$, Eq.~\eqref{eq:MCcoeff};
\item we subtract $c\, d(x)$ directly from $\Delta P_{gg}^{\rm NLL}$;
\item we finally compute $\Delta P_{gq}^{\rm NLL}$ and $\Delta P_{qq}^{\rm NLL}$ according to Eq.~\eqref{eq:xNLLevolmatrix}.
\end{itemize}
The four $\Delta P_{ij}$ constructed in this way are the primary output of the code \texttt{HELL}.

\subsection{Numerical implementation and results}
\label{sec:gammaResults}

\begin{figure*}[t]
  \centering
  \includegraphics[width=0.495\textwidth,page=1]{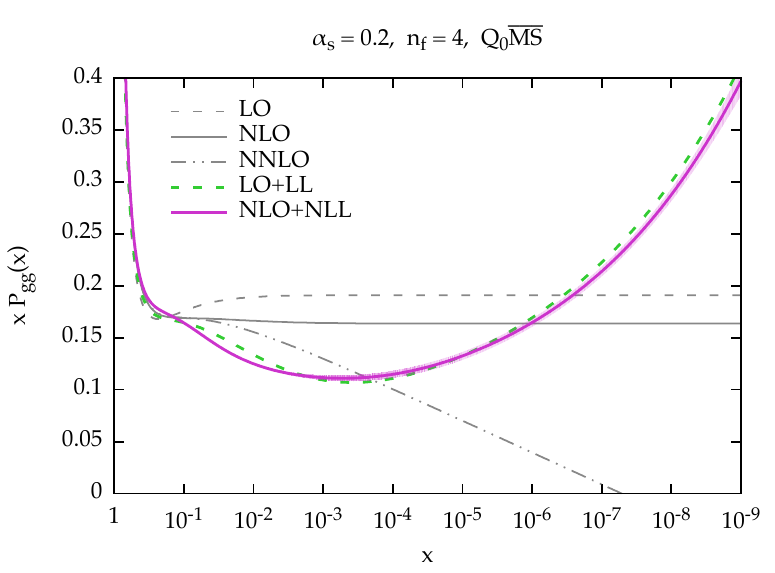}
  \includegraphics[width=0.495\textwidth,page=2]{images/plot_P_nf4_paper.pdf}\\
  \includegraphics[width=0.495\textwidth,page=3]{images/plot_P_nf4_paper.pdf}
  \includegraphics[width=0.495\textwidth,page=4]{images/plot_P_nf4_paper.pdf}
  \caption{\new{The resummed and matched splitting functions at LO+LL (dashed green) and NLO+NLL (solid purple) accuracy:
    $P_{gg}$ (upper left), $P_{gq}$ (upper right), $P_{qg}$ (lower left) and $P_{qq}$ (lower right).
    The fixed-order results at LO (dashed) NLO (solid) and NNLO (dot-dot-dashed) are also shown (in black).
    The NLO+NLL result also includes an uncertainty band, as described in the text.
    The plots are for $\as=0.2$ and $n_f=4$ in the $Q_0\MSbar$ scheme.}}
  \label{fig:Pres}
\end{figure*}

The numerical implementation of the resummation of $\gamma_+$ is quite
challenging.  The main difficulty comes from the fact that several
ingredients of the resummation procedure are not available in a closed
analytic form, but they are only defined as zeroes of implicit
equations which must be solved numerically in the complex plane.
Moreover, these equations can depend on functions which are themselves
computed as zeros of implicit equations (see
App.~\ref{app:gammaplusresdetails} for more details and explicit examples).
While for real $N$ one can rely on robust root-finding algorithms
such as bracketing methods, in the complex plane one must rely on
root-polishing methods whose convergence heavily depends on the
accuracy of the initial guess supplied to the algorithm.  Moreover,
several functions have more than a single branch which satisfy the
zero criterium, hence it is crucial to consistently identify the
correct one.

\begin{figure*}[t]
  \centering
  \includegraphics[width=0.495\textwidth,page=5]{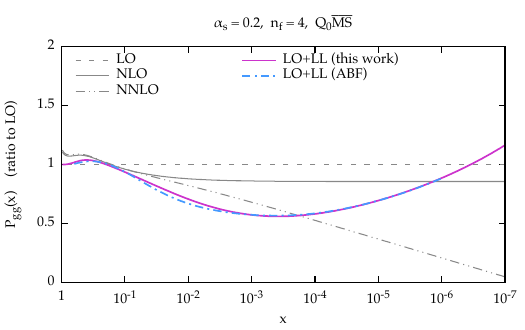}
  \includegraphics[width=0.495\textwidth,page=6]{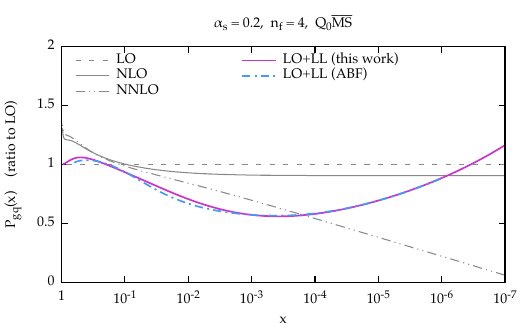}
  \caption{\new{Ratio of fixed-order and resummed LO+LL splitting functions over their LO counterparts, for $P_{gg}$ (left) and $P_{gq}$.
    For comparison, the resummed results of Ref.~\cite{Altarelli:2008aj} are also shown (dot-dashed cyan).
    The plots are for $\as=0.2$ and $n_f=4$. Note at this accuracy the factorization schemes $Q_0\MSbar$ and $\MSbar$ coincide.}}
  \label{fig:PratioLO}
\end{figure*}
\begin{figure*}[t]
  \centering
  \includegraphics[width=0.495\textwidth,page=1]{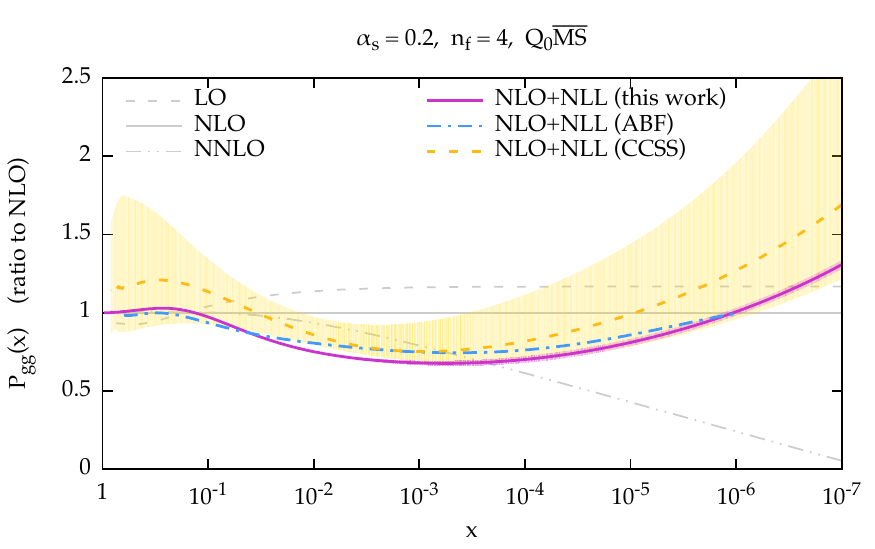}
  \includegraphics[width=0.495\textwidth,page=2]{images/plot_P_nf4_paper_ratio.pdf}\\
  \includegraphics[width=0.495\textwidth,page=3]{images/plot_P_nf4_paper_ratio.pdf}
  \includegraphics[width=0.495\textwidth,page=4]{images/plot_P_nf4_paper_ratio.pdf}
  \caption{\new{Ratio of fixed-order and resummed NLO+NLL splitting functions over their NLO counterparts.
    The plots are for $\as=0.2$ and $n_f=4$ in the $Q_0\MSbar$ scheme, except for CCSS curve, which uses a different factorization scheme.}}
  \label{fig:PratioNLO}
\end{figure*}

We circumvent the above difficulties by computing $\gamma_+^{\rm (N)LO+(N)LL}(N,\as)$ only along the contour for
Mellin inversion, which we parametrize, in the upper plane $\Im N>0$ (in the lower plane we use the complex conjugate path),
as $N=c+t\exp\frac{i3\pi}2$, where $t\in[0,\infty)$ is the integration variable and $c\sim1$ is a parameter whose value
is adjusted for each value of $\as$ to give optimal convergence properties for the Mellin inversion.
For $t=0$, $N=c$ is real, and we can therefore use robust bracketing root-finiding algorithms
which are guaranteed to converge.
As we move from $N=c$ into the complex plane ($t>0$), we resort to the secant method,
whose reliability entirely depends on our
ability to provide an accurate guess of the root to be found. Our
strategy here consists in proceeding by small steps in $t$,
using for initial guess at each step the value of the function at the previous step.
If the step is fine enough and the function sufficiently well behaved,
this method works well and also avoids jumps across different branches.
Very rarely, when this method fails, we can also use a slower but more stable
minimum-finding algorithm, by turning the problem of finding a zero of a function
into the one of finding the minimum of the absolute value of the function itself.
As a consistency check, we verify that at large $|N|$ (large $t$)
the resummed expression becomes asymptotically close to the known fixed-order result.

Using this strategy, we construct tables of values of
$\Delta\gamma_+^{\rm (N)LL}(N,\as)$ along the contour for a grid in
$\as$, one grid for each value of $n_f=3,4,5,6$.  The tables also
contain information about the leading singularities of $\gamma_+$,
namely the position of the leading poles and value of their residues.
We keep the code which produces the tables private, and use the tables
as primary ingredients for the public code presented in this work.

The public code \texttt{HELL} reads the provided tables as input files, and performs the remaining steps for the resummation.
In particular, it constructs the resummed quark anomalous dimension $\Delta\gamma_{qg}^{\rm NLL}(N,\as)$
according to the procedure described in Sect.~\ref{sec:resgammaqg}, along the Mellin inversion contour.
It then performs the inverse Mellin transform and reconstruct the full singlet splitting function matrix,
as described in Sect.~\ref{sec:resmatrix}.
(A similar strategy is used for the resummed coefficient functions, see Sect.~\ref{sec:coeff-func}.)

The \texttt{HELL} code, while being quite flexible and numerically stable, is rather ``heavy'' ($\sim100$~MB)
due to the size of the files which contain the tabulated $\Delta\gamma_+^{\rm (N)LL}$, and also slow
due to the presence of numerical integration
(although we implemented a dynamical caching which speeds up multiple evaluation in a single run).
Therefore, we created a higher-level variant of the code, dubbed \texttt{HELL-x},
which reads pre-tabulated (with \texttt{HELL}) splitting functions (and coefficient functions) on a $\{\as,x\}$ grid for each value of $n_f$ and interpolates them.
Flexibility is lost but this version is very light (a few MB) and very fast.
\texttt{HELL-x} has been interfaced to the evolution code \texttt{APFEL}~\cite{Bertone:2013vaa},
and will be in future used to obtain high-energy resummed PDF fits.

\new{We now present some representative results for the resummed splitting functions for $\as=0.2$ and $n_f=4$.
In Fig.~\ref{fig:Pres} we show all four entries of the evolution matrix:
$P_{gg}$ (upper-left panel) and $P_{gq}$ (upper-right panel), $P_{qg}$ (lower-left panel) and $P_{qq}$ (lower-right panel).
The values of $x$ range from $1$ to $10^{-9}$.
 We include in the plots the fixed-order splitting functions at LO (dashed), NLO (solid) and NNLO (dot-dot-dashed) in black.
At resummed level, we show in solid purple the NLO+NLL result, while the LO+LL result is shown in dashed green
and is present only for $P_{gg}$ and $P_{gq}$, as the other two entries do not have any leading logarithmic enhancement.
At NLO+NLL we have to specify the factorization scheme. 
As previously mentioned, we adopt $Q_0\MSbar$, which is convenient from an all-order viewpoint.
We recall that the difference between $\MSbar$ and $Q_0\MSbar$ starts relative order $\Ord(\as^3)$
and therefore the fixed-order splitting functions start to differ only beyond NNLO.}
We see that at large $x$ the resummation has no effect, due to the damping, so the resummed result smoothly
matches onto the fixed order. At smaller $x$, the resummed result grows. The effect is more pronounced in the case of $P_{qg}$,
where the growth starts immediately, while for $P_{gg}$ the growth is delayed by an initial decrease,
a well-known feature of subleading small-$x$ contributions~\cite{Ciafaloni:2003rd,Altarelli:2005ni,White:2006yh}.
Similarly, we see the same effect on $P_{qq}$ and $P_{gq}$,
where the contribution of the resummation is just $C_F/C_A$ times the contribution on the left plots, Eq.~\eqref{eq:xNLLevolmatrix}.
\new{As far as $P_{gg}$ and $P_{gq}$ are concerned, we observe a nice perturbative convergence of the resummed and matched results,
with the NLO+NLL being a very small correction over the LO+LL, especially when compared with the fixed-order perturbative behaviour at small $x$.
This convergence derives from the stability of the resummation of $\gamma_+$, mostly determined by the 
the constraints imposed by symmetrization and momentum conservation, as described in Sect.~\ref{sec:resgamma+}.}

\new{We have included in Fig.~\ref{fig:Pres} an ``uncertainty band'' for the NLO+NLL result.
This band is determined by replacing in Eq.~\eqref{eq:recursion} $\dot\gamma$ with $-\as\beta_0\gamma$.
As Eq.~\eqref{eq:recursion} is derived under the assumption of linearity of $\gamma_+^{\rm LL^\prime}$,
both expressions are equally valid, and the difference between the two can be taken as a measure of the uncertainty coming
from subleading corrections beyond the linear approximation.
The distance between our default construction and this alternative approach is then symmetrized, thus giving the band.
We acknowledge that the resulting uncertainty is just one of the many sources of uncertainties of the resummation, as coming from
the various approximations described before and from subleading terms.
However, we think that the uncertainty shown in Fig.~\ref{fig:Pres} is a good representative
of the uncertainty from subleading contributions.
We have indeed verified that other variations of subleading terms, e.g., the actual form of $\gamma_+^{\rm LO,sing}$ in Eq.~\eqref{eq:gammaLLp},
leads to similar effects.
On the other hand, the uncertainty on the resummation of $\gamma_+$ is likely to be much smaller, due to the many constraints
on its construction, as confirmed the agreement between different groups~\cite{Dittmar:2009ii}, as well as by the good convergence of the gluon entries.
Clearly, the overall uncertainty from all sources of ambiguities will be larger, but we believe the shape and the relative size among the various entries
is likely to be well represented by the current band.
}

\new{
We now move to the comparison of our results with other approaches. 
To better highlight the impact of the resummation, we show the comparisons in terms of ratios over the fixed-order splitting functions.
In Fig.~\ref{fig:PratioLO} the ratio of resummed LO+LL splitting functions over the LO ones are presented for $P_{gg}$ and $P_{gq}$
(at this order, only the gluon components are affected by resummation).
Along with our curves, the ABF results of Ref.~\cite{Altarelli:2008aj} are also shown in dot-dashed cyan
(the plotted range is limited in $x$ due to the available information from the original paper).
The fixed NLO (solid) and NNLO (dot-dot-dashed) are also shown (in gray) for comparison's sake.
Overall, we observe good agreement with our result.
The tiny deviation is due to a different treatment of the $n_f$ dependence of the result,
see App.~\ref{app:gammaplusresdetails} for more detail.
Interestingly, we observe that at large $x$ the resummed results tend to follow the shape of the NLO and NNLO results,
before merging onto the LO due to the damping, perhaps an indication that higher order contributions predicted by the
resummation go in the right direction even far from the small-$x$ region.
Note also that the LO+LL ratio is basically identical for $P_{gg}$ and $P_{gq}$, a small difference being visible only
at large $x$. This is easily understood by noting that the small-$x$ behaviour of both fixed-order
and resummed results are simply related by a color factor $C_F/C_A$.
}

\new{The comparison of the NLO+NLL resummed results are shown in Fig.~\ref{fig:PratioNLO}.
Here, not only we compare our results to the ones obtained by ABF in Ref.~\cite{Altarelli:2008aj} but also to the resummed splitting function calculated in Ref.~\cite{Ciafaloni:2007gf} (henceforth the CCSS approach).
The latter also comes with a (yellow) uncertainty band which is obtained from renormalization scale variation.
While the agreement with ABF is still rather good, there are more significant deviations,
especially in the quark entries, which come from many sources.
For $P_{qg}$ (and $P_{qq}$), we use the LL$^\prime$ anomalous dimension, Eq.~\eqref{eq:gammaLLp},
while ABF used the full NLL anomalous dimension. Moreover, we implement differently the large-$N$ subtraction,
as discussed in Sect.~\ref{sec:resgammaqg}, and we also have different numerical implementations,
as we adopt a Borel-Pad\'e summation for the series Eq.~\eqref{eq:gamma_qg_ss_rc}.
These differences also affect $P_{gg}$ (and $P_{gq}$), due to Eq.~\eqref{eq:gamma_gg} but their numerical impact appears to be smaller.
Note that for these gluon splitting functions we also have differences at large $x$ due to our implementation
of momentum conservation, Eq.~\eqref{eq:gamma_gg_momcons}.
Unfortunately, our simple uncertainty band does not fully cover all these differences, especially at larger $x$.
When comparing to CCSS, we see that the gluon entries $P_{gg}$ and $P_{gq}$ are in decent agreement,
our result lying at the lower edge of the CCSS band.
The quark entries $P_{qg}$ and $P_{qq}$, however, are quite different both in shape and in size.
It is clear that these entries are affected by larger uncertainties, as demonstrated by both our and the CCSS bands,
as well as by the large perturbative corrections in the fixed order.
Therefore, it is likely that such a difference is a manifestation of this ambiguity, which could be fixed
only by a NNLL computation.
Note also that the CCSS results are obtained in a scheme which is not exactly the $Q_0\MSbar$,
and it is well known that differences between schemes can be significant at resummed level
(see e.g.\ the comparison of the $\MSbar$ and $Q_0\MSbar$ in Ref.~\cite{Altarelli:2008aj}).
}

\section{Resummation of perturbative coefficient functions}
\label{sec:coeff-func}

We now turn our attention to the resummation of small-$x$ enhanced
contribution to collinearly factorized partonic coefficient
functions.
The general formalism for the resummation of inclusive cross sections is based on $k_t$-factorization,
which was derived a long time ago~\cite{Catani:1990xk,Catani:1990eg,Collins:1991ty,Catani:1993ww,Catani:1993rn,Catani:1994sq}
and it is known to LL\footnote{Here by LL we mean the lowest non-trivial logarithmic order,
  which is sometimes NLL in absolute order counting, as in the case of DIS discussed later in this section.}
for an increasing number of cross sections and
distributions~\cite{Ball:2001pq,Hautmann:2002tu,Marzani:2008az,Harlander:2009my,Marzani:2008uh,Diana:2010ef,Caola:2011wq,Caola:2010kv,Forte:2015gve}.

The ABF approach for resumming coefficient functions was developed in Ref.~\cite{Ball:2007ra} and applied to the case of DIS structure functions in Ref.~\cite{Altarelli:2008aj}.
The crucial point to note is that, analogously to the case of PDF evolution, the resummation of formally subleading running coupling corrections plays a crucial role. 
The procedure that we will describe in this section does take these effects into account but departs from the original ABF method in that the resummation is performed directly in transverse momentum space rather than in Mellin moment space. Although the two procedures are formally equivalent, as we shall discuss below, the momentum-space technique significantly helps with two shortcomings of the Mellin-space approach. First of all, computing Mellin moments of $k_t$-factorized cross sections with respect of the gluons' $\kb$ often constitutes the bottle-neck of a calculation.
Secondly, running coupling corrections in Mellin space are included order-by-order in perturbation theory and then a Borel summation of the resulting series is performed, resulting in potential numerical instabilities.
Thus, working directly in transverse-momentum space avoids dealing with asymptotic series and opens up the possibility of performing resummed calculations for processes for which Mellin moments cannot be computed analytically.


In order to keep the notation simple, we consider a process
with only one hadron in the initial state, such as DIS. The generalization to two hadronic legs is
straightforward, as discussed in Ref.~\cite{Marzani:2008uh}.
Because we are interested in the high-energy limit, we limit ourselves
to consider the singlet sector. 
Although we work in transverse-momentum space, we find convenient to take Mellin moments with respect the longitudinal momentum fractions and to work with cross sections in $N$ space.
The generic cross section is then given by (henceforth we use $\as=\asq$)
\begin{align} \label{eq:coll_fact}
\sigma(N,Q^2)
  & = C_g\(N,\as\) f_g (N,Q^2)  + C_q\(N,\as\) f_q (N,Q^2) \nonumber\\
   =& \, C_+\(N,\as\) f_+ (N,Q^2)  + C_-\(N,\as\) f_- (N,Q^2),
\end{align}
where $f_q$ and $f_g$ are the quark-singlet and gluon PDFs, respectively,
and in the second line we have transformed to the basis of the eigenvectors of singlet DGLAP evolution.
In DIS, $\sigma$ can be either the structure function $F_2$ or $F_L$ ($F_3$ is non-singlet),
up to a normalization factor (for precise definitions, see Ref.~\cite{Catani:1994sq}).
Since only $f_+(N,Q^2)$ resums at small $x$, we have a single coefficient
function, $C_+(N,\as)$, which is affected by small $x$ enhancements.
We will come back later in Sect.~\ref{sec:rotation} on the precise definition of $C_+$ in terms of $C_g$ and $C_q$.

It is known, e.g.~\cite{Catani:1990xk,Catani:1990eg,Collins:1991ty}, that in the high-energy limit a different,
more general, form of factorization holds, even away from the collinear limit\footnote{For a more general discussion on transverse-momentum dependent factorization, we refer the Reader to Ref.~\cite{Collins:1350496}.}:
\beq \label{eq:kt_fact}
\sigma(N,Q^2) = \int d\kb^2 \,
{\cal C}\(N,\frac{\kb^2}{Q^2},\asq\)\,{\cal F}_g(N,\kb^2)
\eeq
where ${\cal F}_g(N,\kb^2)$ is the unintegrated ($k_t$ dependent) gluon PDF and ${\cal C}(N,\kb^2/Q^2,\as)$ is the off-shell coefficient function,
i.e.\ the coefficient function for the partonic process with an off-shell initial state gluon.

In the high-energy limit, the unintegrated gluon density can be related to the standard resummed PDF
\beq\label{eq:Udef}
{\cal F}_g(N,\kb^2) = \U\(N,\frac{\kb^2}{Q^2}\)\, f_+(N,Q^2).
\eeq
Before discussing the form of $\U(N,\kb^2/Q^2)$, we immediately
observe that once the relation Eq.~\eqref{eq:Udef} between the integrated and unintegrated PDFs is established,
by comparing the gluon contribution in $k_t$-factorization Eq.~\eqref{eq:kt_fact}
and the high-energy contribution in collinear factorization Eq.~\eqref{eq:coll_fact} we are able to write
\beq \label{eq:res_C+}
C_+\(N,\as\) = \int d\kb^2 \, {\cal C}\(N,\frac{\kb^2}{Q^2},\as\)\, \U\(N,\frac{\kb^2}{Q^2}\).
\eeq
This equation represents our main formula for the implementation of high-energy resummation
in the coefficient functions. In \texttt{HELL}, the $\kb^2$ integral is evaluated numerically,
given the off-shell cross section ${\cal C}(N,\kb^2/Q^2,\as)$ in $\kb$ space,
and an actual form of $\U(N,\kb^2/Q^2)$, which will be discussed in the next subsection.
Note that LL accuracy only requires to calculate ${\cal C}$ to lowest order in $\as$. Moreover, its $N$ dependence is also subleading and one can set $N=0$.

\subsection{The evolution factor}

We now turn to discussing the form of $\U(N,\kb^2/Q^2)$ in Eq.~\eqref{eq:Udef}.
As clear from Eq.~\eqref{eq:Udef}, it first evolves the largest eigenvector PDF from $Q^2$ to
the scale $\kb^2$, where it then converts it to the unintegrated gluon PDFs.
It can be understood either in terms of the all-order gluon Green's
function~\cite{Catani:1990xk,Catani:1990eg,Collins:1991ty,Catani:1994sq}
or as the evolution kernel of a generalized ladder expansion~\cite{Caola:2010kv}.
At lowest order and fixed coupling, the form of $\U$ is known~\cite{Catani:1994sq}
\beq \label{eq:U_LL}
\U_s\(N,\frac{\kb^2}{Q^2}\) = \mathcal{R}(\gamma_s) \frac{d}{d\kb^2}\(\frac{\kb^2}{Q^2}\)^{\gamma_s},
\eeq
where $\gamma_s$ is the anomalous dimension obtained from the leading order BFKL kernel with duality at fixed coupling $\as=\asq$.
We also note the scheme-dependent factor $\mathcal{R}(\gamma_s)$ that originates from the correct treatment of collinear singularities, the calculation of which requires a more accurate analysis away from $d=4$ space-time dimensions.
In the commonly used $\MSbar$ scheme this factor reads~\cite{Catani:1994sq}
\begin{align} \label{eq:MSB}
\mathcal{R}_{\MSbar}(M)&=\sqrt{\frac{-1}{M}\frac{\Gamma\(1-M\) \chi_0 \(M\)}{\Gamma\(1+M\) \chi_0^\prime \(M\)} }
\nonumber \\ & \times \exp \left \{M \psi(1)+\int_0^{M} d c \, \frac{\psi^\prime(1)-\psi^\prime \(1-c\)}{\chi_0(c)} \right\},
\nonumber \\
&= 1+ \Ord \(M^3\),
\end{align}
where $\chi_0(M)$ is the eigenvalue of the leading-order BFKL kernel and $\Gamma(x)$ and $\psi(x)$ are the Euler gamma and di-gamma functions, respectively.
In $Q_0 \MSbar$ instead we simply have
\beq \label{eq:Q0MSB}
\mathcal{R}_{Q_0\MSbar}(M)=1.
\eeq
\new{Comparing the last line of Eq.~(\ref{eq:MSB}) to Eq.~(\ref{eq:Q0MSB}) we see that the difference between the two schemes starts at relative $\Ord\( \as^3\)$. }
It is also useful to write the scheme-dependent factor as 
\beq
\mathcal{R}(M) = \int_0^\infty d \xi \, \xi^{M-1} \bar{\mathcal{R}}(\xi);
\eeq
while it is not straightforward to find a closed analytic form of $\bar{\mathcal{R}}_{\MSbar}(\xi)$
from Eq.~\eqref{eq:MSB}, in the $Q_0\MSbar$ scheme we simply have $\bar{\mathcal{R}}_{Q_0 \MSbar}(\xi)=\delta(1-\xi)$.

The running-coupling generalization of Eq~\eqref{eq:U_LL} that we implement is
\begin{align} \label{eq:U_LL_RC}
\U\(N,\frac{\kb^2}{Q^2}\) &= 
\int_0^\infty \frac{d q_2^2}{q^2_2} \bar{\mathcal{R}}\(\frac{q_2^2}{\kb^2} \) \nonumber \\ &\times
\exp \[  \int_{\kb^2}^{q_2^2} \frac{d q_1^2}{q_1^2} \gamma_+(N,\as(q_1^2)) \] \nonumber \\ &\times
\frac{d}{d \kb^2}
\exp \[  \int_{Q^2}^{\kb^2} \frac{d q_1^2}{q_1^2} \gamma_+(N,\as(q_1^2)) \],
\end{align}
where $\gamma_+$ is the resummed anomalous dimension.
Note that by substituting $\gamma_+ \to \gamma_s$, at fixed
coupling, we recover the lowest-order result Eq.~\eqref{eq:U_LL}. The
general structure of the result appears fairly complicated because of
the presence of the scheme factor $\bar{\mathcal{R}}$. However, in the preferred
scheme $Q_0 \MSbar$ the first two lines of Eq.~\eqref{eq:U_LL_RC}
evaluate to unity and the result simplifies to 
\beq \label{eq:U_LL_RC_Q0}
\U_{Q_0\MSbar}\(N,\frac{\kb^2}{Q^2}\) = \frac{d}{d \kb^2}
\exp \[  \int_{Q^2}^{\kb^2} \frac{d q_1^2}{q_1^2} \gamma_+(N,\as(q_1^2)) \],
\eeq
where we recognize the derivative of a DGLAP evolution factor.

\subsection{Basis transformation and collinear subtraction}
\label{sec:rotation}

Once $C_+$ is computed according to Eq.~\eqref{eq:res_C+}, one can use the relation between
$C_+$ and $C_g,C_q$ to obtain resummed expressions for $C_g$ and $C_q$.
This relation can be trivially obtained from the transformation matrix that diagonalizes the
DGLAP evolution equation in the singlet sector, Eq.~\eqref{eq:Rdef}, leading to
\beq
C_\pm = C_g + r_\pm C_q.
\eeq
At fixed coupling, as discussed in Sect.~\ref{sec:resgamma+},
diagonalizing the evolution equation simply amounts to diagonalizing the singlet anomalous dimension matrix,
and using Eq.~\eqref{eq:r+-defFC} would lead to the simple relations
\begin{align}\label{eq:rotatedCFsFC}
 C_+ = C_g + \frac{\gamma_{qg}}{\gamma_+-\gamma_{qq}} C_q, \nonumber\\
 C_- = C_g + \frac{\gamma_{qg}}{\gamma_--\gamma_{qq}} C_q.
\end{align}
However, as previously discussed, when running coupling effects are taken into account, a transformation that diagonalizes the evolution matrix does not in general diagonalize the evolution equation, since the derivative with respect to $Q^2$ acts on the transformation matrix, Eq.~\eqref{eq:DGLAPdiag}, producing an additional contribution which is in general not diagonal.
Furthermore, we note that in contrast to the case of $\gamma_{gg}$, here a more careful treatment of these running coupling corrections is required in order to guarantee the all-order cancellation of collinear singularities that may be present in $C_+$.

Finding the general transformation matrix that diagonalizes the singlet evolution equation
is not an easy task.
However, because our goal is to find a running coupling version of Eq.~\eqref{eq:rotatedCFsFC},
a full solution is not needed, as long as we limit ourselves to the LL accuracy. 

To this purpose, it is convenient to consider the logarithmic derivative of Eq.~\eqref{eq:coll_fact} with respect to $Q^2$
\begin{align} \label{eq:coll_fact_der}
\frac{d\sigma(N,Q^2)}{d\ln Q^2}
  &= \(\frac{dC_g}{d\ln Q^2} + C_g\gamma_{gg} + C_q\gamma_{qg}\) f_g(N,Q^2) \nonumber\\&+ \(\frac{dC_q}{d\ln Q^2} + C_q\gamma_{qq} + C_g\gamma_{gq}\) f_q (N,Q^2) \nonumber\\
  &= \(\frac{dC_+}{d\ln Q^2} + C_+\gamma_+\) f_+ (N,Q^2) \nonumber\\&+ \(\frac{dC_-}{d\ln Q^2} + C_-\gamma_-\) f_- (N,Q^2).
\end{align}
The first two and last two lines of Eq.~\eqref{eq:coll_fact_der} are related by the same
transformation matrix that relates first and second line of Eq.~\eqref{eq:coll_fact}.
However, the logarithmic derivative already produces running coupling contributions, making further running-coupling effects
on the transformation matrix genuinely subleading.
Thanks to this observation, we can use the fixed-coupling transformation matrix to relate the various terms in Eq.~\eqref{eq:coll_fact_der}.
For the $+$ component we are mostly interested into, this leads to the equation
\begin{align}
\frac{dC_+}{d\ln Q^2} + C_+\gamma_+ &= \frac{dC_g}{d\ln Q^2} + C_g\gamma_{gg} + C_q\gamma_{qg}
\nonumber \\+& \frac{\gamma_{qg}}{\gamma_+-\gamma_{qq}}\(\frac{dC_q}{d\ln Q^2} + C_q\gamma_{qq} + C_g\gamma_{gq}\).
\end{align}
We now need to understand the logarithmic order of each contribution, and keep only those terms which are LL.
First, we observe that the logarithmic derivative of the coefficient function is one logarithmic order higher
than the coefficient function itself. This suggest that all derivative terms could be thrown away, leading back Eq.~\eqref{eq:coll_fact}.
However, the key point of the resummation of running coupling effects is exactly to keep those subleading terms
which are suppressed by $\as\beta_0$, which are precisely those coming from these derivatives.
Next, from the analysis of the previous section, we know that $\gamma_{gg}$ and $\gamma_{gq}$ are LL, while $\gamma_{qg}$ and $\gamma_{qq}$ are NLL.
Since \emph{to all orders} $C_q$ is of the same logarithmic order as $C_g$ (as we shall see later in this section),
this suggests that only the first two terms on the right-hand side of Eq.~\eqref{eq:coll_fact_der} should be kept.
However, some of those terms can be leading if there is a \emph{fixed-order} contribution
in the coefficient function which is of higher logarithmic order than the coefficient function itself.
This is for instance the case of the DIS structure function $F_2$: in this case,
both $C_g$ and $C_q$ are NLL (in absolute counting), but the fixed-order expansion of $C_q$ is $C_q=1+\Ord(\as)$,
where the first $\Ord(\as^0)$ term is formally LL. When this is the case, the term $C_q\gamma_{qg}$
with $C_q$ replaced by its fixed-order superleading contribution leads to a leading contribution to the equation and must be retained. Finally, the last contribution is genuinely subleading.

After all these consideration, and further approximating $\gamma_{gg}$ with $\gamma_+$ (the difference being subleading),
we end up with the equation
\beq\label{eq:rot_eq}
\frac{dC_+}{d\ln Q^2} + C_+\gamma_+ = \frac{dC_g}{d\ln Q^2} + C_g\gamma_+ + C_q\gamma_{qg},
\eeq
which can be easily solved introducing an exponential factor
\beq\label{eq:Uevol}
U\(N,\frac{Q^2}{Q_0^2}\) = \exp \[  \int_{Q_0^2}^{Q^2} \frac{d \mu^2}{\mu^2} \gamma_+(N,\as(\mu^2)) \],
\eeq
so that Eq.~\eqref{eq:rot_eq} becomes
\beq
\frac{d}{d\ln Q^2}\(UC_+\) = \frac{d}{d\ln Q^2}\(UC_g\) + C_q\gamma_{qg}.
\eeq
The solution is then
\beq\label{eq:rotatedCFs}
C_+(N,\as) = C_g(N,\as) + C_q(N,\as) \, U_{qg}(N,Q^2)
\eeq
having defined
\begin{align}\label{eq:collsubRC}
U_{qg}(N,Q^2) &=
\int_{Q_0^2}^{Q^2} \frac{dq^2}{q^2} \gamma_{qg}(N,\as(q^2)) \nonumber\\ &\times \exp \[  \int_{Q^2}^{q^2} \frac{d\mu^2}{\mu^2} \gamma_+(N,\as(\mu^2)) \],
\end{align}
where $Q_0$ is the scale at which $U$ vanishes (which is the position of the Landau pole),
and we have left $C_q$ outside the integral because it is either $1$ or $0$.
Eq.~\eqref{eq:rotatedCFs} represents the running coupling version of the first of Eq.~\eqref{eq:rotatedCFsFC},
at LL. As a cross check, we can easily verify that if the coupling does not run we get
\beq\label{eq:rotatedCFsFCLL}
C_+(N,\as) = C_g(N,\as) + C_q(N,\as)\frac{\gamma_{qg}(N,\as)}{\gamma_+(N,\as)},
\eeq
which is indeed equivalent to Eq.~\eqref{eq:rotatedCFsFC} up to subleading terms.
With similar arguments, it is also possible to show that the solution in presence of running
of the equation for $C_-$ leads exactly to its fixed-coupling counterpart, second line of Eq.~\eqref{eq:rotatedCFsFC},
up to NLL terms.
Note that this suggests that the generalization of the transformation matrix $R$, Eq.~\eqref{eq:Rdef},
is simply obtained by using (up to subleading corrections)
\beq\label{eq:r+}
r_+ = U_{qg}(N,Q^2)
\eeq
and the fixed-coupling value of $r_-$.

We have now all the ingredients to obtain resummed expressions for $C_g$ and $C_q$.
From Eq.~\eqref{eq:rotatedCFs} we immediately have
\begin{align}\label{eq:res_Cg}
C_g \(N,\as\) &= C_+\(N,\as\) - C_q \(N,\as\) U_{qg}(N,Q^2) \nonumber\\
&= \int d\kb^2 \, {\cal C}\(N,\frac{\kb^2}{Q^2},\as\)\, \U\(N,\frac{\kb^2}{Q^2}\) \nonumber\\&
  - C_q (N,\as) U_{qg}(N,Q^2),
\end{align}
where in the second line we have used Eq.~\eqref{eq:res_C+}.
As we already discussed, the $C_q$ subtraction is suppressed by a NLL term,
so this term is present only when $C_q$ has a fixed-order contribution which is superleading.
This is the case of the DIS structure function $F_2$, where $C_{2,q}$ is NLL (in absolute counting) and $C_{2,q}=1+\Ord(\as)$.
In this case, we can just replace $C_q$ with $1$ and get
\begin{align}\label{eq:res_C2}
C_{2,g} \(N,\as\) &= \int d\kb^2 \, {\cal C}_2\(N,\frac{\kb^2}{Q^2},\as\)\, \U\(N,\frac{\kb^2}{Q^2}\) \nonumber\\&- U_{qg}(N,Q^2).
\end{align}
In other cases, such as the longitudinal structure functions $F_L$, $C_{L,q}$ is still NLL in absolute counting but does not contain
any superleading fixed-order contributions, as it starts at $\Ord(\as)$; therefore, the $C_q$
contribution is genuinely subleading and one finds
\beq\label{eq:res_CL}
C_{L,g} \(N,\as\) = \int d\kb^2 \, {\cal C}_L\(N,\frac{\kb^2}{Q^2},\as\)\, \U\(N,\frac{\kb^2}{Q^2}\).
\eeq
The resummed expressions for $C_q$ can be found from the second of Eq.~\eqref{eq:rotatedCFsFC},
\begin{align}\label{eq:res_Cq1}
C_q(N,\as)
&= \frac{\gamma_{qq}(N,\as)-\gamma_-(N,\as)}{\gamma_{qg}(N,\as)} \times \nonumber\\& \times \Big[C_g(N,\as)-C_-(N,\as)\Big].
\end{align}
We note that $C_-$ does not contain any logarithmic enhancements to all orders and therefore it can be safely evaluated at fixed-order (NLO) and at $N=0$. Furthermore, we can make use of the high-energy color-charge relation Eq.~\eqref{eq:gamma_qq} and arrive at
\begin{align}\label{eq:res_Cq}
C_q(N,\as)
&= \frac{C_F}{C_A} \Big[C_g(N,\as)-C_-(0,\as)\Big],
\end{align}
which shows that $C_q$ and $C_g$ are of the same logarithmic order, as anticipated.

In order to perform the matching to the fixed-order, we find useful to introduce ($i=g,q$)
\begin{align}\label{eq:deltaC}
\Delta_n C_{i}(N,\as)&= C_{i}(N,\as)-\sum_{k=0}^n \as^k C^{(k)}_{i}(N),
\end{align}
where $C^{(k)}_{i}(N)$ is the $k$-th order coefficient of the $\as$ expansion of the resummed result $C_{i}(N,\as)$.
Hence, the second contribution subtracts from the first term (the resummed result) its expansion
up to the perturbative order we want to match to, e.g. $n=1$ is NLO, $n=2$ is NNLO.
In this notation the resummed and matched contribution is simply given by
\begin{align}\label{eq:matchedC}
 C^\text{N$^n$LO+LL}_{i}(N,\as)&=C^\text{N$^n$LO}_{i}(N,\as)+\Delta_n C_{i}(N,\as),
\end{align}
where the resummed contributions $\Delta_n C_{i}(N,\as)$ are computed by \texttt{HELL}, while the fixed-order parts have to be provided by an external code.
Note that the color-charge relation Eq.~\eqref{eq:res_Cq} reduces to
\beq\label{eq:deltaCcolorcharge}
\Delta_n C_{q}(N,\as) = \frac{C_F}{C_A} \Delta_n C_{g}(N,\as)
\eeq
when written in terms of $\Delta_n$ contributions, provided $n\geq1$.
Note also that these $\Delta_nC_{q}$, with $n\geq1$, can be seen as the resummed contributions to the pure-singlet
quark coefficient functions~\cite{Catani:1994sq,Marzani:2008uh}.

\subsection{Equivalence between transverse-momentum space and Mellin space resummations}
\label{sec:ABFcomparison}

In this section we want to compare our result Eq.~\eqref{eq:res_C+} with running coupling Eq.~\eqref{eq:U_LL_RC}
to the analogous result obtained in the ABF approach~\cite{Altarelli:2008aj}, which is performed in Mellin space.
For convenience, and without loss of generality, we work in the $Q_0\MSbar$ scheme, and using the definition Eq.~\eqref{eq:Uevol}
we can write
\begin{align}
\U_{Q_0\MSbar}\(N,\frac{\kb^2}{Q^2}\)
&= \frac{d}{d\kb^2} U\(N,\frac{\kb^2}{Q^2}\) = \frac{1}{Q^2} \frac{d}{d\xi} U\(N,\xi\),
\end{align}
where we introduced the dimensionless variable $\xi=\kb^2/Q^2$.
We can thus write Eq.~\eqref{eq:res_C+} as
\begin{align}\label{eq:integral1}
C_+\(N,\as\)
  &=  \int d\xi \, {\cal C}\(N,\xi,\as\)\, \frac{d}{d\xi}U\(N,\xi\). 
\end{align}
In the ABF approach the resummation of coefficient functions closely follows the one of the quark anomalous dimension $\gamma_{qg}$,
where in place of the function $h(M)$, Eq.~\eqref{eq:hqgser}, the Mellin transform of the off-shell coefficient function
with respect to $\kt$ is used.
Therefore we define the so-called impact factor,\footnote{In the literature, this is usually called $h$.}
\begin{align}
\tilde{\cal C}(N,M,\as) 
&= M \int_0^\infty d\xi\, \xi^{M-1} {\cal C}\(N,\xi,\as\),
\end{align}
where $\tilde{\cal C}(N,M,\as) $ admits an expansion in powers of $M$
\beq\label{eq:CMexp}
\tilde{\cal C}(N,M,\as) = \sum_{k} \tilde{\cal C}_k(N,\as) M^k.
\eeq
Note that $k\geq-1$ for processes that are not two-particle irreducible in the high-energy limit
and therefore their lowest-order diagrams exhibit a collinear singularity,
as in the case of $F_2$, while $k\geq0$ for processes without such collinear singularity, as in the case of $F_L$.
The inverse Mellin transform is given by
\begin{align}\label{eq:CinvMell2}
{\cal C}\(N,\xi,\as\)
&= \int_{c-i\infty}^{c+i\infty} \frac{dM}{2\pi i} \,\xi^{-M} \frac1M \tilde{\cal C}(N,M,\as) \nonumber\\&=
\sum_{k}\tilde{\cal C}_k(N,\as) \int_{c-i\infty}^{c+i\infty} \frac{dM}{2\pi i} \xi^{-M} M^{k-1} \nonumber \\
&= \[\tilde{\cal C}_{-1}(N,\as) \ln\frac1\xi
  + \tilde{\cal C}_{0}(N,\as)\] \theta(1-\xi) \nonumber\\ &\quad
  + \sum_{k\geq1}\tilde{\cal C}_k(N,\as) \[\partial_\nu^{k-1} \delta(\nu-\ln\xi)\]_{\nu=0},
\end{align}
where the integration contour is a standard Mellin inversion contour, with $0<c<1$.
The resummed expression for the coefficient function $C_+$ can be now found substituting Eq.~\eqref{eq:CinvMell2} into Eq.~\eqref{eq:integral1}. The integral over $\xi$ can be performed in all cases and we find
\begin{align}
C_+\(N,\as\)
&= \tilde{\cal C}_{-1}(N,\as) \[\ln\xi_0\, U(N,\xi_0) + \int_{\xi_0}^1 \frac{d\xi}\xi U\(N,\xi\)\]
\nonumber\\ &\quad + \tilde{\cal C}_{0}(N,\as) \[1 - U(N,\xi_0) \]\nonumber\\
&\quad+ \sum_{k\geq1} \tilde{\cal C}_k(N,\as) \[\partial_\nu^k U\(N,e^\nu\)\]_{\nu=0}, 
\label{eq:connectiontoABF1}
\end{align}
where we have introduced a lower integration limit $\xi_0$. This lower limit is equal to $0$ in the fixed coupling
case, but in the running coupling case we have
\beq\label{eq:lowerlimit}
\xi_0 = \exp\[-\frac1{\as\beta_0}\]
\eeq
due to the presence of the Landau pole.
Note that, assuming $\gamma_+>0$ (as appropriate close to the pole), $U(N,\xi_0)=0$ so Eq.~\eqref{eq:connectiontoABF1} simplifies
\begin{align}
C_+\(N,\as\)
&= \tilde{\cal C}_{-1}(N,\as) \int_{\xi_0}^1 \frac{d\xi}\xi U\(N,\xi\)
\nonumber\\&+ \sum_{k\geq0} \tilde{\cal C}_k(N,\as) \[\partial_\nu^k U\(N,e^\nu\)\]_{\nu=0},
\label{eq:connectiontoABF}
\end{align}
which represents an equivalent form of Eq.~\eqref{eq:res_C+}.

Let us now focus on the simpler case without collinear singularities, $\tilde{\cal C}_{-1}=0$.
We want to show that the sum in Eq.~\eqref{eq:connectiontoABF} corresponds
to the procedure adopted in ABF, under some assumptions on the form of the resummed anomalous dimension.
In particular, we recover ABF assuming that the dominant running coupling effects
are determined by 1-loop running of the lowest power of $\as$
appearing in the anomalous dimension. In other words, one makes the approximation
(as usual $\as=\asq$)
\beq\label{eq:gamma_approx}
\gamma_+\(N,\as(\mu^2)\) = \frac{\gamma_+(N,\as)}{1+\as\beta_0\ln(\mu^2/Q^2)},
\eeq
which is an exact expression at LO, where $\gamma_+\(N,\as(\mu^2)\) = \as(\mu^2) \gamma^{(0)}_+(N)$.
In order to better describe the exact anomalous dimension which is not
simply linear in $\as$, one can replace
\beq\label{eq:asb0_replacement}
\as\beta_0 \to 
-\frac{\dot \gamma_+(N,\as)}{\gamma_+(N,\as)}
=\frac{\as^2\beta_0}{\gamma_+(N,\as)}\frac{d}{d\as}\gamma_+(N,\as),
\eeq
so that the $\mu^2$ derivative of Eq.~\eqref{eq:gamma_approx} in $\mu^2=Q^2$ is correct (and the 1-loop structure is kept).
In this particular approximation, the $\nu$-derivatives in Eq.~\eqref{eq:connectiontoABF} satisfy the recursion
\begin{align}
&\[\partial_\nu^{k+1} U_{\rm ABF}\(N,e^\nu\)\]_{\nu=0} \nonumber\\&
= \[\partial_\nu^{k} U_{\rm ABF}\(N,e^\nu\)\]_{\nu=0} \[\gamma_+(N,\as) - k\as\beta_0\] \nonumber\\
&= \[\partial_\nu^{k} U_{\rm ABF}\(N,e^\nu\)\]_{\nu=0} \[\gamma_+(N,\as) + k \frac{\dot\gamma_+(N,\as)}{\gamma_+(N,\as)}\],
\label{eq:[gamma]}
\end{align}
where $U_{\rm ABF}$ indicates the evolution factor Eq.~\eqref{eq:Uevol} computed with $\gamma_+$ from Eq.~\eqref{eq:gamma_approx}.
We recognize the recursion defined in Eq.~\eqref{eq:recursion}.
This recursive construction is exactly the method employed by ABF to perform
the running coupling resummation of coefficient functions.
Therefore, we recover the ABF result\footnote
{At small $N$. We treat the large $N$ behaviour differently, see discussion in App.~\ref{app:gammaqgcomparison}.}
(in the case of no collinear singularities, as in $F_L$)
\beq\label{eq:CL_ABF}
C_{L,g} = \sum_{k\geq0} \tilde{\cal C}_{L,k}(0,\as) \[\partial_\nu^k U_{\rm ABF}\(N,e^\nu\)\]_{\nu=0},
\eeq
where we further computed the expansion coefficients $\tilde{\cal C}_k$ in $N=0$.
However, we recall that the resulting series is divergent, and cannot be summed analytically,
so sophisticated numerical techniques with limited numerical accuracy are needed
in order to use Eq.~\eqref{eq:connectiontoABF}, see App.~\ref{app:Borel}.

In presence of a collinear singularity, the first term in Eq.~\eqref{eq:connectiontoABF}
proportional to $\tilde{\cal C}_{-1}(N,\as) $ does not vanish. Additionally,
the collinear subtraction due to $C_q$ must be included.
In the ABF approach, the subtraction is written first in Mellin space as
$\as h(M)/M$, with $h(M)$ defined in Sect.~\ref{sec:resgammaqg},
and subtracted directly at the level of inverse Mellin integrand, leading to
(in the case of $F_2$)
\beq\label{eq:res_C2_ABF}
C_{2,g} = \sum_{k\geq0}\[\tilde{\cal C}_{2,k}(0,\as) - \as h_{k+1}\] \[\partial_\nu^k U_{\rm ABF}\(N,e^\nu\)\]_{\nu=0},
\eeq
where $h_k$ are the expansion coefficients of $h(M)$ in powers of $M$, Eq.~\eqref{eq:hqgser},
and the collinear term $\tilde{\cal C}_{2,-1}(0,\as)/M$ cancels against the first term $-h_0/M$
of the collinear subtraction, since $\tilde{\cal C}_{2,-1}(0,\as)=\as h_0$.
In our approach, Eq.~\eqref{eq:res_Cg} together with Eq.~\eqref{eq:connectiontoABF} leads to
\begin{align}\label{eq:res_C2_}
C_{2,g} &= \sum_{k\geq0}\tilde{\cal C}_{2,k}(N,\as) \[\partial_\nu^k U\(N,e^\nu\)\]_{\nu=0} \nonumber\\
&\quad+ \tilde{\cal C}_{2,-1}(N,\as) \int_{\xi_0}^1 \frac{d\xi}\xi \, U\(N,\xi\) - U_{qg}(N,Q^2) \nonumber\\
&= \sum_{k\geq0}\tilde{\cal C}_{2,k}(N,\as) \[\partial_\nu^k U\(N,e^\nu\)\]_{\nu=0} \nonumber\\
&\quad + \int_{\xi_0}^1 \frac{d\xi}\xi \, U\(N,\xi\) \[\as h_0 - \gamma_{qg}(N,\as(Q^2\xi))\].
\end{align}
To prove the equivalence of Eq.~\eqref{eq:res_C2_ABF} and \eqref{eq:res_C2_} under the ABF assumptions
we need to express Eq.~\eqref{eq:gamma_qg_ss_rc} with the help of Eq.~\eqref{eq:[gamma]} as
\beq\label{eq:gamma_qg_ABF}
\gamma_{qg}(N,\as(Q^2\xi)) = \as \sum_{k\geq0} h_{k} \[\partial_\nu^k U_{\rm ABF}\(N,e^\nu\)\]_{\nu=\ln\xi}.
\eeq
Plugging this into Eq.~\eqref{eq:res_C2_} it is immediate to verify that the $h_0$ term cancels,
and the integral can be computed to lead to exactly Eq.~\eqref{eq:res_C2_ABF}.
Note that the usage of the running-coupling version of the basis transformation discussed
in Sect.~\ref{sec:rotation} is crucial to obtain the correct result.
Had one used the fixed-coupling version, the collinear singularity would not cancel.

\begin{figure}[t]
  \centering
  \includegraphics[width=0.45\textwidth,page=6]{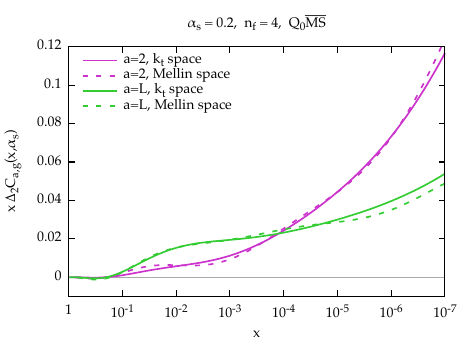}
  \caption{The resummation contribution $\Delta_2 C_{a,g}$ computed with Mellin-space (dashed) and $k_t$-space (solid) resummation for both $a=2$ and $a=L$,
    using $\as=0.2$ and $n_f=4$ in the $Q_0\MSbar$ scheme.}
  \label{fig:comparison}
\end{figure}
Therefore, we have shown that our transverse momentum space derivation and the Mellin space resummation adopted by ABF
are completely equivalent, even though the current result is more general and does not rely on the assumptions of
Eq.~\eqref{eq:gamma_approx} and \eqref{eq:asb0_replacement}.
A numerical comparison of Mellin-space and $k_t$-space resummation is performed in Fig.~\ref{fig:comparison}.
The plot shows $\Delta_2 C_{a,g}$ Eq.~(\ref{eq:deltaC}) for both $a=2,L$, with $\as=0.2$ and $n_f=4$ \new{in the $Q_0\MSbar$ scheme}.
We observe that the two approaches give indeed the same result. We note however that the Mellin space implementation
suffers from numerical instabilities, which determine small oscillations around the actual result.
These oscillations become more severe at larger $\as$, and disappear at smaller $\as$.
We note that these numerical instabilities are related to the approximate Borel-Pad\'e method used for the Mellin space
implementation, which necessarily uses a limited amount of information
(i.e., a finite number of coefficients of the $M=0$ expansion, see App.~\ref{app:Borel}).
In Ref.~\cite{Altarelli:2008aj} a different ``truncated'' Borel method was used,
which did not develop oscillation; however, also in that case the amount of information used was limited,
while in our $\kt$-space approach we make use of all the information residing in the off-shell cross section.

\subsection{Numerical implementation and results}
\label{sec:cfnumeric}

We now turn to the numerical implementation of the resummation of coefficient functions in \texttt{HELL}.
Starting from Eq.~\eqref{eq:res_C+} written as in Eq.~\eqref{eq:integral1}, we integrate by parts
(the boundary terms vanish at $\xi\to\infty$ thanks to $\cal C$ and in $\xi_0$ thanks to $U$)
and evaluate the off-shell cross section at $N=0$ (since its $N$ dependence is subleading),
\begin{align}\label{eq:res_C+_ibp}
C_+\(N,\as\) &= -\int_{\xi_0}^\infty d\xi \, \frac{d}{d\xi}{\cal C}\(0,\xi,\as\) \, U(N,\xi).
\end{align}
As the resummation of coefficient functions is at present accurate only at LL,
we may conveniently compute $U(N,\xi)$ using the LL$^\prime$ anomalous dimension introduced in Eq.~\eqref{eq:gammaLLp},
\begin{align}\label{eq:ULLp}
U(N,\xi) &= \exp\int_1^\xi \frac{d\zeta}{\zeta}\gamma_+^{\rm LL'}\(N,\as(Q^2\zeta)\).
\end{align}
However, since $\as$ in the evolution factor is evaluated at $Q^2\zeta$ with $\zeta$ ranging up to $\xi$,
and $\xi$ is integrated over all accessible values, the resummed anomalous dimension
should be computed at extreme values of $\as$, from $0$ to $\infty$.
This is problematic in practice, since the resummed anomalous dimension is itself computed numerically
as described in Sect.~\ref{sec:anom-dim}, and it is numerically challenging to reach both
high and low values of $\as$.

Therefore, a convenient implementation consists in adopting the approximation 
Eq.~\eqref{eq:gamma_approx}, possibly together with the replacement Eq.~\eqref{eq:asb0_replacement},
as in ABF. Under this assumption, the integral in the exponent can be computed analytically,
and we have
\beq\label{eq:UABF}
U_{\rm ABF}(N,\xi) = \Big(1+r(N,\as)\ln\xi\Big)^{\gamma_+^{\rm LL'}(N,\as)/r(N,\as)}
\eeq
with
\beq\label{eq:rdef}
r(N,\as) = -\frac{\dot\gamma_+^{\rm LL'} (N,\as)}{\gamma_+^{\rm LL'} (N,\as)}.
\eeq
This expression is advantageous because the integral in the evolution
factor has been computed analytically and it only requires $\gamma_+^{\rm LL'}$ and its $\as$ derivative
at $\as=\asq$.

\begin{figure*}[t]
\centering
  \includegraphics[width=0.495\textwidth,page=3]{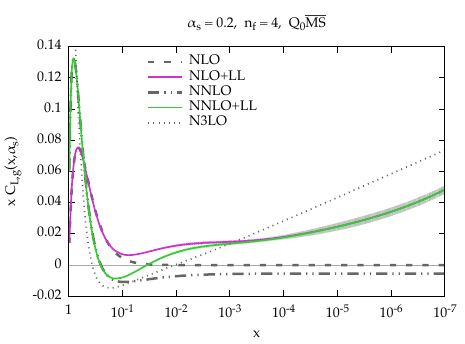}
  \includegraphics[width=0.495\textwidth,page=4]{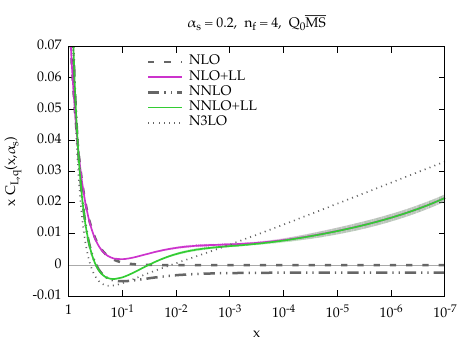}
  \caption{\new{The resummed and matched coefficient function $C_{L,i}$ at NLO+LL accuracy (solid purple) and at NNLO+LL accuracy (solid green).
    The gluon case $i=g$ is on the left-hand panel, the quark-singlet case $i=q$ is on the right-hand panel.
    The fixed-order results are also shown in black: NLO in dashed, NNLO in dot-dot-dashed and N$^3$LO in dotted.
    Our result also includes an uncertainty band, as described in the text.
    The plots are for $\as=0.2$ and $n_f=4$ in the $Q_0\MSbar$ scheme.}}
  \label{fig:cL}
\end{figure*}
\begin{figure*}[t]
  \centering
  \includegraphics[width=0.495\textwidth,page=1]{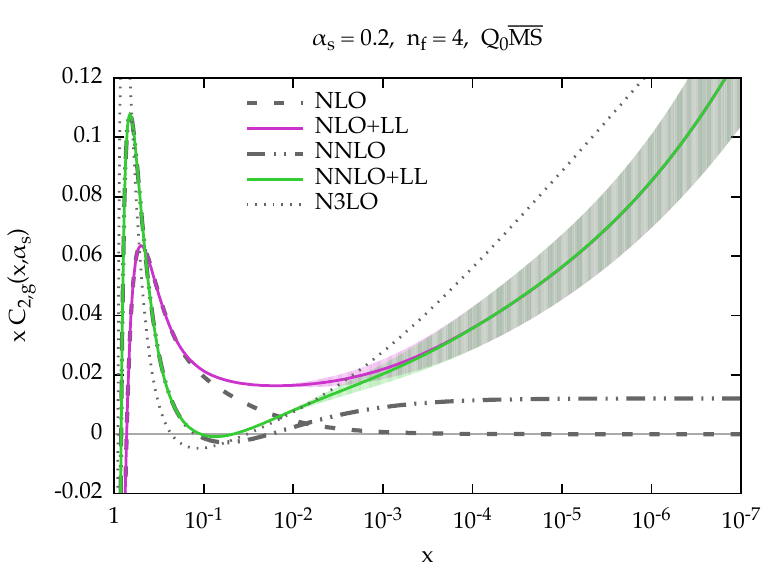}
  \includegraphics[width=0.495\textwidth,page=2]{images/plot_C_nf4_paper.pdf}\\
  \caption{Same as Fig.~\ref{fig:cL}, but for the coefficient functions $C_{2,i}$.}
  \label{fig:c2}
\end{figure*}

We now turn to the specific case of massless DIS. For an observable without collinear singularity,
such as the longitudinal structure function, we simply have
\begin{align}\label{eq:res_CLfinal}
C_{L,g}\(N,\as\) &= -\int_{\xi_0}^\infty d\xi \, \frac{d}{d\xi}{\cal C}_L\(0,\xi,\as\) \, U_{\rm ABF}(N,\xi).
\end{align}
For processes with collinear singularities, we further need the collinear subtraction $U_{qg}$,
Eq.~\eqref{eq:collsubRC}, to obtain $C_g$, Eq.~\eqref{eq:res_Cg}.
Computing the integral Eq.~\eqref{eq:collsubRC} numerically, even within the approximation Eq.~\eqref{eq:gamma_approx},
is challenging due to the need of integrating $\gamma_{qg}$ over a range of $\as$ from $\asq$ to $\infty$.
In principle, we could find an approximation similar to Eq.~\eqref{eq:gamma_approx} for $\gamma_{qg}$.
However, we propose here a different approach, based on the ABF formulation Eq.~\eqref{eq:gamma_qg_ABF},
which allows us to write
\begin{align}\label{eq:Uqg}
U_{qg}(N,Q^2) &= \as\sum_{k\geq0} h_{k+1} \[\partial_\nu^k U_{\rm ABF}\(N,e^\nu\)\]_{\nu=0} \nonumber \\ &\quad+ \as h_0 \int_{\xi_0}^1 \frac{d\xi}\xi \, U_{\rm ABF}\(N,\xi\).
\end{align}
The sum in Eq.~\eqref{eq:Uqg} can be computed as we compute $\gamma_{qg}$ itself.
In fact, the computation is identical, except that the $h_k$ coefficients are all shifted by a unity.
This way, we can pre-tabulate it once for all, 
and use it for any observable with collinear singularities.
The integral term in Eq.~\eqref{eq:Uqg} can be combined with the integral in Eq.~\eqref{eq:res_C+_ibp},
so that the collinear subtraction is performed at the level of the integrand,
leading to a more reliable numerical implementation.
So, for $C_2$, we have finally
\begin{align}\label{eq:C2resfinal}
C_{2,g}\(N,\as\)
= -\int_{\xi_0}^\infty d\xi \[\frac{d}{d\xi}{\cal C}_2\(0,\xi,\as\) + \frac{\as h_0}\xi \theta(1-\xi)\] \nonumber \\ \times  U_{\rm ABF}\(N,\xi\)  -\as\sum_{k\geq0} h_{k+1} \[\partial_\nu^k U_{\rm ABF}\(N,e^\nu\)\]_{\nu=0}.
\end{align}
From these resummed expressions, we can then construct the resummed contributions, $\Delta_nC_{g}(N,\as)$, Eq.~\eqref{eq:deltaC}
(see also App.~\ref{app:disexpansion}), and $\Delta_nC_q(N,\as)$ from Eq.~\eqref{eq:deltaCcolorcharge}.
At this point, as we did for the splitting functions, we damp the resummed contributions in $x$ space
multiplying by $(1-x)^2$ to ensure a smooth matching onto the fixed order.\footnote
{In practice, a smoother matching to NLO is obtained if $\Delta_1$ is derived from $\Delta_2$, as detailed in App.~\ref{app:dis}.}

The resummed and matched partonic coefficient functions are shown in Fig.~\ref{fig:cL}
in the case of $C_{L}$, and in Fig.~\ref{fig:c2} in the case of $C_2$.
In both cases, the gluonic coefficient functions are shown on the left-hand panel,
while the quark ones on the right-hand panel.
The solid purple line is for NLO+LL, while the solid green for NNLO+LL. The resummation is performed in $Q_0 \MSbar$.
\new{Analogously to the case of the splitting functions, the size of the uncertainty band is obtained from the symmetrized difference between the calculation performed with $r$ as given in Eq.~\eqref{eq:rdef} or its linearized version $r=\as\beta_0$.
The corresponding fixed-order results are also shown: NLO in dashed, NNLO in dot-dot-dashed and N$^3$LO~\cite{Vermaseren:2005qc} in dotted.
The plots are for $\as=0.2$ and $n_f=4$.
}

\begin{figure}[t]
  \centering
  \includegraphics[width=0.45\textwidth,page=5]{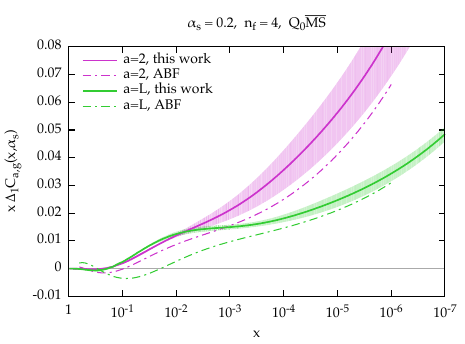}
  \caption{\new{Comparison of the resummation contribution $\Delta_1 C_{a,g}$ as obtained in this work (solid)
    versus the ABF results (dot-dashed) of Ref.~\cite{Altarelli:2008aj}
    for both $a=2$ (purple) and $a=L$ (green), using $\as=0.2$ and $n_f=4$ in the $Q_0\MSbar$ scheme.}}
  \label{fig:comparisonABF}
\end{figure}

\new{The comparison to the ABF approach is done in Fig.~\ref{fig:comparisonABF}, where the resummed contribution $\Delta_1 C_{a,g}$ ($a=2,L$) is shown.
We note that our results are in general agreement with the ones of the ABF paper~\cite{Altarelli:2008aj},
especially if we focus on the longitudinal coefficient functions $C_{L,i}$, $i=g,q$.
In the case of $C_{2,i}$, differences are instead more pronounced.
This should not come as a surprise because, as discussed at length,
the resummation for the coefficient functions differs by various subleading terms.
We stress once again that we have verified (see e.g.\ Fig.~\ref{fig:comparison}) that the resummation performed in Mellin space (as in Ref.~\cite{Altarelli:2008aj})
gives identical results (modulo numerical instabilities at large $\as$)
as our $k_t$-space formulation, as long as the same $\gamma_+$ is used and the
same subtraction of the large-$N$ terms is adopted.
Therefore, the difference comes from both the different way of subtracting the large-$N$ behaviour (see discussion in App.~\ref{app:gammaqgcomparison})
and the fact that we use $\gamma_+^{\rm LL'}$ rather than $\gamma_+^{\rm NLL}$.
Moreover, note that the band is indeed larger in the $C_2$ case, confirming that subleading effects in $C_2$ are more pronounced than in $C_L$.
}
In particular, a direct comparison with the expressions of Ref.~\cite{Altarelli:2008aj}
shows that our result differs by constant terms at $\Ord(\as)$ and $\Ord(\as^2)$ in the resummed $\gamma_+$,
which lead to formally NLL and NNLL differences in the resummed coefficient functions.
We conclude that, in absence of a well motivated preference for these subleading contributions,
both results have to be considered as equally valid at the present logarithmic accuracy,
and the ambiguity can only be fixed by computing (resumming) the NLL contributions in the coefficient functions.
\new{At larger $x$, we observe a significant deviation between our result and ABF for $C_L$,
which is not well represented by the band. In this case the difference has to do with the large-$N$ matching,
and we expect our matching procedure to perform better than ABF.
}

\section{Conclusions and outlook} \label{sec:conclusions}

In this paper we have discussed the resummation of high-energy, i.e.\ small-$x$, logarithms that affect both the evolution of collinearly-factorized parton densities and perturbative coefficient functions. Despite a wealth of calculations have been performed in $k_t$-factorization, the framework that allows for high-energy resummation, very few phenomenological studies that incorporate both fixed-order and resummed calculations existed, essentially because of the complexity of the running-coupling resummation of the DGLAP and BFKL evolution kernels.

In this paper we have overcome this obstacle and we have developed a computer code named  \texttt{HELL} (High Energy Large Logarithms), available for download at
$$\href{http://www.ge.infn.it/~bonvini/hell}{\texttt{www.ge.infn.it/\textasciitilde{}bonvini/hell}} \,,$$
that enables one to obtain small-$x$ resummed DGLAP splitting and partonic coefficient functions. The code is based on the formalism developed by Altarelli, Ball and Forte (ABF), with several improvements that avoid numerical instabilities and facilitate the future inclusion of different processes. The main innovation with respect to the ABF original procedure consists in performing the resummation of perturbative coefficient functions from the off-shell cross section in transverse-momentum space rather than in Mellin-moment conjugate space. Therefore, partonic off-shell cross sections computed in $k_t$-factorization can be directly used, without the necessity of performing Mellin transformations with respect to the initial-state gluons' virtualities, which is often the bottle-neck of this kind of calculations.

\new{We have provided resummed results for the splitting functions in the singlet sector, both at LO+LL and NLO+NLL and, as a proof of principle, we have also performed the resummation for the massless DIS structure functions $F_2$ and $F_L$, at NLO+LL and NNLO+LL.
We have provided a qualitative estimate of the theoretical uncertainty by varying subleading contributions that are related to the running of the strong coupling. We have found that this uncertainty is rather small for the gluon splitting functions $P_{gg}$ and $P_{gq}$, essentially because their resummation is mostly driven by the all-order behaviour of the leading eigenvalue in the singlet sector, which is under good theoretical control. On the other hand, the uncertainty is larger for the quark splitting functions $P_{qg}$ and $P_{qq}$, as well as for the closely-related DIS coefficient functions, for which we only control the first tower of logarithmic contributions. This feature also appears in the comparisons to ABF and CCSS. Indeed, all the approaches considered here are in decent agreement for the gluon splitting functions, while they significantly differ in the quark sector, which is also plagued by rather large uncertainties.}

We see this, rather technical, paper as the first necessary step towards a rich program of small-$x$ phenomenology. First, we would like to use the results presented here to perform a PDF fit of DIS data that consistently include small-$x$ resummation in both parton evolution and perturbative coefficient functions, especially in view of the recent final release of HERA data~\cite{Abramowicz:2015mha}. 
These small-$x$ resummed PDF fits will be performed in the NNPDF global analysis framework~\cite{Ball:2014uwa} and preliminary results have been presented in~\cite{Rojo:2016kwu}.

Furthermore, having at hand resummed PDFs, we will perform a study of small-$x$ effects at high-energy hadron colliders, such as the LHC or an FCC. In particular at FCC, because of the extremely large center-of-mass energy, low-$x$ effects in processes like Higgs or vector boson production are expected to become very important. In this respect, the study of electro-weak boson production via the Drell-Yan mechanism offers an almost unique environment to look for deviation from standard DGLAP dynamics.
Finally, some of us have recently developed frameworks to combine small-$x$ resummation with threshold~\cite{Ball:2013bra} and transverse-momentum resummation~\cite{Marzani:2015oyb} and we look forward to performing phenomenological studies of joint resummation in the context of Higgs and electro-weak bosons productions.

\section*{Acknowledgments}
We are deeply indebted to Richard Ball and Stefano Forte for introducing us to small-$x$ physics and for their constant guidance and encouragement over the many years of work on this topic.
\new{We thank Gavin Salam for providing us with the results of the CCSS approach}.
We have also benefited from many useful discussions with Fabrizio Caola, Giovanni Ridolfi and Juan Rojo. 
The work of M.B.\ is supported by an European Research Council Starting Grant ``PDF\ 4BSM: Parton Distributions in the Higgs Boson Era''.
The work of S.M.\ is supported in part by the U.S.\ National Science Foundation, under grant PHY-0969510, the LHC Theory Initiative.
The work of T.P.\ is supported by an STFC Rutherford Grant ST/M004104/1.

\appendix
\section{Details on the resummation of $\gamma_+$}
\label{app:gammaplusresdetails}

In this appendix we give further details about the resummation of
$\gamma_+$.  A comprehensive treatment of this topic can be found in
Refs.~\cite{Altarelli:2005ni,Altarelli:2008aj} and it is beyond the
purpose of this paper.  Here, we collect the relevant formulae needed
for the numerical implementation of our version of the ABF resummation
procedure, pointing out the changes and improvements we introduced with respect
to the literature.

\subsection{Double-leading contributions and symmetrization}
\label{app:double-lead-contr}

As we briefly explained in Sect.~\ref{sec:resgamma+}, one of the
ingredients for building a stable DL expansion of the BFKL kernel
(and by duality of the resummed DGLAP anomalous dimension) is
\emph{symmetrization}~\cite{Salam:1998tj}, i.e.\ the construction of  a kernel which satisfies
symmetry properties otherwise spoiled by subleading terms.  As
explained e.g.\ in Refs.~\cite{Salam:1998tj,Altarelli:2005ni}, the kernel $\chi$ in the
fixed coupling limit satisfies $\chi(M,\as)=\chi(1-M,\as)$
if the kinematic is symmetric upon exchange of the virtualities $Q^2$ and $k^2$.
This is e.g.\ true for gluon-gluon
scattering where the kinematic is $x=\sqrt{Q^2 k^2}/s$, but the
symmetry is broken for DIS-like kinematics where $x\sim Q^2/s$.  If
$\chi_\sigma$ and $\chi_\Sigma$ are the kernels obtained with a
symmetric (gluon-gluon scattering) and asymmetric (DIS) choice of $x$
respectively, one can however show that the following (equivalent)
relations hold
\begin{align}
  \label{eq:chisymmetry}
  \chi_\Sigma\Big(M+\frac{1}{2}\, \chi_\sigma(M,\as),\as\Big) = \chi_\sigma(M,\as), \nonumber \\
  \chi_\sigma\Big(M-\frac{1}{2}\, \chi_\Sigma(M,\as),\as\Big) = \chi_\Sigma(M,\as).
\end{align}
In the ABF approach one constructs, at a given logarithmic accuracy,
a symmetric kernel $\chi_\sigma$ such that
\begin{equation} \label{eq:chisymmetric}
  \chi_\sigma(M,\as)=\chi_\sigma(1-M,\as)
\end{equation}
and a corresponding asymmetric kernel $\chi_\Sigma$ satisfying
Eq.~\eqref{eq:chisymmetry}, by means of the introduction of so-called
\emph{off-shell kernels}.  An off-shell kernel is a kernel
$\chi(M,N,\as)$ which depends on both $M$ and $N$ and is related
to its \emph{on-shell} counterpart $\chi(M,\as)$ and to the dual
anomalous dimension $\gamma(N,\as)$ by the equation
\begin{equation}
  \label{eq:onshellness}
  \chi(M,N,\as) = N,
\end{equation}
evaluated at
\begin{equation}
  \label{eq:Neqchi}
  N=\chi(M,\as)
\end{equation}
or
\begin{equation}
  \label{eq:Meqgamma}
  M=\gamma(N,\as)
\end{equation}
respectively.  With the expression ``putting on-shell'' an
off-shell kernel, we mean solving Eq.~\eqref{eq:onshellness} for
$\chi(M,\as)$ while imposing Eq.~\eqref{eq:Neqchi} or solving it for
or $\gamma(N,\as)$ while imposing Eq.~\eqref{eq:Meqgamma}.  The
resulting on-shell kernel and anomalous dimension satisfy the duality
relation in Eq.~\eqref{eq:duality} by construction.  The solutions to
these equations, similarly to the duality equations, must be found via
numerical methods in the complex plane.

The procedure for the construction of the off-shell kernel is based on
the separation of collinear $M\leq 0$ and anti-collinear $M\geq 1$
singularities in the original expression of $\chi$, i.e.\ rewriting
\begin{equation}
  \chi(M,\as)  = \chi_+(M,\as) + \chi_-(M,\as)
\end{equation}
where the collinear piece $\chi_+$ has poles in $M=0,-1,-2,\ldots$ and
the anti-collinear piece $\chi_-$ has poles in $M=1,2,\ldots$.  As
suggested by the symmetry relation in Eq.~\eqref{eq:chisymmetric}, we
can define the two contributions such that they satisfy
$\chi_-(M,\as) =\chi_+(1-M,\as)$.  The two off-shell kernels are thus
roughly defined as
\begin{align}
  \chi_\Sigma(M,N,\as) \approx {}& \chi_+(M,\as) + \chi_+(1-M+N,\as) \\
  \chi_\sigma(M,N,\as) \approx {}& \chi_+\left(M+\frac{N}{2},\as\right) \nonumber \\&+ \chi_+\left(1-M+\frac{N}{2},\as\right).
\end{align}
The actual expressions for the kernels are more involved and take into
account several technical details which are thoroughly discussed in
Ref.~\cite{Altarelli:2005ni}.  In the following we collect explicit
formulae for the results.

The anomalous dimension $\gamma_+^{\Sigma,\text{LO}}(N,\as)$ in
Eq.~\eqref{eq:gammaLOrc} is obtained by putting on-shell the kernel
\begin{align} \label{eq:chiLO}
  \chi_\Sigma^{\text{LO}}(M,N,\as)&=\chi_s\left(\frac{\as}{M}\right)+\chi_s\left(\frac{\as}{1-M+N}\right)
   \nonumber \\&+\as\tilde\chi_0(M,N)+\chi_{\text{mom}}(N,\as),
\end{align}
where here and in the following $\chi_{\text{mom}}$ is a subleading
contribution which implements momentum conservation.  This can be
taken to be of the form
\begin{equation}
  \chi_{\text{mom}}(N,\as) = c_m f_{\text{mom}}(N), \label{eq:chimom}
\end{equation}
where $f(0) = f(\infty)=0$ and $f_m(1)=1$, e.g.
\begin{equation}
  f_{\text{mom}}(N) = \frac{4 N}{(1+N)^2},
\end{equation}
and $c_m$ is such that the final kernel satisfies momentum
conservation
\begin{equation}
  \chi_{\Sigma}(0,1,\as) = 1.
\end{equation}
In Eq.~\eqref{eq:chiLO}, $\chi_s$ represents the dual of the LO DGLAP
anomalous dimension and it is defined by the implicit equation
\begin{equation} \label{eq:dualityLO}
  \gamma_+^{\text{LO}}\(\chi_s\(\frac{\as}{M}\),\as\) \equiv   \as \gamma_+^{(0)}\left(\chi_s\left(\frac{\as}{M}\right)\right) = M.
\end{equation}
The kernel $\tilde\chi_0(M,N)$ contains the BFKL LL contributions
and is constructed from the LO BFKL kernel
\begin{equation} \label{eq:chi0}
  \chi_0(M)=\frac{C_A}{\pi}\big(2\psi(1)-\psi(M)-\psi(1-M)\big),
\end{equation}
by performing the off-shell extension discussed above and removing the
double counting with the DGLAP contributions.  Its expression reads
\begin{align} \label{eq:chi0bar}
  \tilde\chi_0\left(M,N\right)  &= \frac{C_A}{\pi}\Big(\, \psi(1)+\psi(1+N)-\psi(1+M) \nonumber \\ &\qquad\qquad-\psi(2-M+N)\, \Big).
\end{align}

At NLO the term $\gamma_+^{\Sigma,\text{NLO}}$ appearing in
Eq.~\eqref{eq:gammaNLOrc} differs by the fixed-coupling on-shell dual
$\gamma_+^{\Sigma,\text{NLO},\text{fc}}$ of the NLO kernel by a running
coupling correction to the duality relation, according to
\begin{align} \label{eq:gammarcpert} \gamma_+^{\Sigma,\text{NLO}}(N,\as)&=\gamma_+^{\Sigma,\text{NLO},\text{fc}} (N,\as) \nonumber \\-&\beta_0\as\left(\frac{\chi''_0\left(\gamma_s(\as/N)\right)\chi_0\left(\gamma_s(\as/N)\right)}{2\left(\chi'_0\left(\gamma_s(\as/N)\right)\right)^2}-1\right),
\end{align}
where the function $\gamma_s$ is the dual of the LO BFKL kernel
\begin{equation}
  \as\, \chi_0(\gamma_s(\as/N)) = N.
\end{equation}
The anomalous dimension $\gamma_+^{\Sigma,\text{NLO},\text{fc}}$ is obtained by
putting on-shell the kernel
\begin{align} \label{eq:chifinal}
  \chi_\Sigma^{\text{NLO}}(M,N,\as)+\as^2\chi_1^{\beta_0}\(M-\frac{N}{2},N\) \nonumber \\+\beta_0\as^2\left(\frac{\chi_0\left(M,N\right)}{M}-\frac{C_A}{\pi M^2}\right),
\end{align}
where
\begin{align} 
\label{eq:chi0bar}
  \chi_{0}\left(M,N\right)
  &=\tilde\chi_0\left(M,N\right)\\ &\quad + \frac{C_A}{\pi}\(\frac1M+\frac1{1-M+N} \) \nonumber\\
  \chi_1^{\beta_0}(M,N)&=-\beta_0\frac{C_A}{\pi}\psi'\left(2-M+\tfrac{N}{2}\right) \label{eq:chi1b0} \\
  \chi_\Sigma^{\text{NLO}}(M,N,\as) &= \chi_{s,\text{NLO}}\left(M,\as\right)\nn
&\quad+\chi_{s,\text{NLO}}\left(1-M+N,\as\right) \nn
&\quad +\as\tilde\chi_0(M,N)+\as^2\tilde\chi_1(M,N)\nn
&\quad+\chi_{\text{mom}}(N,\as). \label{eq:chiNLO}
\end{align}
In the last equation $\chi_{s,\text{NLO}}$ is the dual of the NLO DGLAP
anomalous dimension
\begin{equation} \label{eq:dualityNLO}
  \gamma_+^{\text{NLO}}(\chi_{s,\text{NLO}}(M,\as),\as) = M.
\end{equation}
Our construction of the kernel $\tilde\chi_1(M,N)$ follows the
procedure outlined in Ref.~\cite{Altarelli:2005ni} (see in particular
Appendix~A of that reference), but the result slightly differs since
we separate the collinear and anti-collinear singularities in the
whole range $-\infty<M<+\infty$ rather than just on a finite interval
(the impact of this on the resummed splitting functions is however
very small and formally of higher twist).  The result can be written
as
\begin{equation} \label{eq:chi1tildeasim} \tilde{\chi}_1(M,N)=\tilde{\chi}_1^{\text{u}}(M,N)-\tilde{\chi}_1^{\text{u}}(0,N)+\tilde{\chi}_1^{\text{u}}(0,0).
\end{equation}
where
\begin{align}
  g_1 ={} & -\frac{13 n_f+10 C_A^2n_f}{36 \pi^2C_A} \label{eq:sxg1} \\
  g_2 ={} & -\frac{11C_A^3+2n_f}{12\pi^2C_A} \label{eq:sxg2} \\
  \tilde{\chi}_1^{\text{u}}(M,N) ={} & \breve{\chi}_1(M,N)-g_1\left(\frac{1}{M}+\frac{1}{1-M+N}\right)\nn
&-g_2\left(\frac{1}{M^2}+\frac{1}{(1-M+N)^2}\right)
\end{align}
and
\begin{align} \label{eq:chi1capasim}
   \breve{\chi}_1(M,N) &=  \chi_1(M,N)\nn
&\quad-\frac{1}{2}  \chi_0(M,N)\, \frac{C_A}{\pi}\Big(2\psi '(1+N)-\psi '(M)\nn
&\qquad\qquad\qquad\quad-\psi '(1-M+N)\Big).
\end{align}
The symmetrized kernel $ \chi_1$ is written in terms of the
function $\phi_L^+$ defined by
\begin{align}
  \phi_{L}^{+}(M)& \equiv
  \int_{0}^{1}d x\; \frac{\text{Li}_2(x)}{x+1}\, x^{M-1}
  = \frac{\pi^{2}}{6}\ln(2) \nn
-&  \sum_{k=1}^{\infty}a_{k}
  \left(\frac{M-1}{M-1 + k} \frac{\pi^{2}}{6} + \frac{k
    \left(\psi(M+k)-\psi(1)\right)}{(M-1 + k)^{2}}\right) \label{eq:phiL+},
\end{align}
as
\begin{align} \label{eq:chi1bar}
  &\chi_1\left(M,N\right) \nn
&= -\frac{1}{2}\beta_0\frac{C_A}{\pi}\left(\frac{\pi^2}{C_A^2} \chi_0(M,N)^2-\psi '(M)-\psi '(1-M+N)\right) \nonumber \\
  &  + \frac{C_A^2}{4\pi^2}\left[\left(\frac{67}{9}-\frac{\pi^2}{3}-\frac{10 n_f}{9 C_A}\right) \left(\psi (1)-\psi(M)\right)+3 \zeta (3) \right. \nonumber \\
    &  +\psi ''(M)+ 4\left(\frac{\pi^2}{24}\left(\psi\left(\tfrac{1}{2}+\tfrac{M}{2}\right)-\psi\left(\tfrac{M}{2}\right)\right)+\phi_L^+(M)\right)\nonumber \\
    &  +\frac{3}{4(1-2M)}\left(\psi'(\tfrac{1}{2}+\tfrac{M}{2})-\psi'(\tfrac{M}{2})+\psi'(\tfrac{1}{4})-\psi'(\tfrac{3}{4})\right) \nonumber \\
    &  +\frac{1}{16}\left(1+\frac{n_f}{C_A^3}\right)(2+3M(1-M)) \nonumber \\
    & \times \left\{ \frac{1}{1-2M}\left(\psi'(\tfrac{1}{2}+\tfrac{M}{2})-\psi'(\tfrac{M}{2})+\psi'(\tfrac{1}{4})-\psi'(\tfrac{3}{4})\right) \right. \nonumber \\
    & +\frac{1}{2(1+2M)}\left(\psi'(\tfrac{1}{2}+\tfrac{M}{2})-\psi'(\tfrac{M}{2})+\psi'(-\tfrac{1}{4})-\psi'(\tfrac{1}{4})\right) \nonumber \\
    &  \left. -\frac{1}{2(3-2M)}\left(\psi'(\tfrac{1}{2}+\tfrac{M}{2})-\psi'(\tfrac{M}{2})+\psi'(\tfrac{3}{4})-\psi'(\tfrac{5}{4})\right)\right\} \nonumber \\
    &  \left. \vphantom{\frac{67}{9}} +(M\leftrightarrow1-M+N) \right].
\end{align}

In the numerical implementation of the resummation procedure for
$n_f\neq 0$, in Eq.~\eqref{eq:dualityLO} and~\eqref{eq:dualityNLO} we
do not use the exact eigenvalue of the DGLAP matrix $\gamma_+$.  The
reason for this is the presence of a branch-cut singularity in the
solution for the eigenvalue equations.  Although this branch-cut
cancels out in results for physical observables, in practice the exact
cancellation is spoiled by subleading terms introduced in the
resummation procedure.  Since the cut is on the right of the leading
small-$N$ singularity, it introduces an unphysical oscillating
behaviour in the splitting functions.  One can however observe that
the whole resummation procedure can be consistently carried out by
replacing $\gamma_+$ with any function which has the same small-$N$
behaviour.  In our approach, we replace $\gamma_+$ with the same
function evaluated in $n_f=0$ plus the $n_f$ dependence of the LL
and NLL contributions only.  As usual, we also add a subleading
term which enforces exact momentum conservation.  The only missing
pieces from the resulting DL expansion are thus the $n_f$-dependent
contributions which are not enhanced at small $N$, but these always
cancel out when taking the difference between resummed and unresummed
result.  The final result for $\Delta \gamma_+^\text{LL}$ and
$\Delta \gamma_+^\text{NLL}$ defined by Eq.~\eqref{delta_gammma_+} is
thus correct at the given logarithmic accuracy in both $\ln (1/x)$ and
$\ln Q^2$ but free of spurious branch cuts.  In
Ref.~\cite{Altarelli:2008aj} a slightly different method was used,
where a rational approximation of the whole $n_f$ dependent part was
used, but we find our minimal approach cleaner and more convenient
(note e.g.\ that by adding too many terms to the approximation one
reconstructs an approximation of the branch cut and thus reproduces the unphysical oscillating
behaviour of the result).  We verified that the difference between the
two approaches is numerically very small (and of course formally
subleading).  A minor subtlety arises when dealing with (subleading
but de facto dominant) running coupling effects, which we discuss in
the next section.

\subsection{Running-coupling contributions}
\label{app:runn-coupl-contr}
\new{The leading small-$N$ singularity of the anomalous dimension is
determined by running coupling corrections which, as already mentioned, determine the small-$x$
asymptotic behaviour of the splitting functions and can be resummed by solving the BFKL evolution equation
for $f_+$~\cite{Thorne:1999sg,Thorne:1999rb,Thorne:2001nr}.
In the ABF approach~\cite{Altarelli:2005ni}, the resummation of the dominant running coupling
contributions is encoded in the so-called \emph{Bateman anomalous dimension}
$\gamma^{B,\rm (N) LL}(N,\as)$ appearing in Eq.~\eqref{eq:gammaLOrc}
and~\eqref{eq:gammaNLOrc} (henceforth generically referred to as
$\gamma^{B}(N,\as)$).  The latter is determined from} the solution of the BFKL evolution equation for
$f_+$, obtained from a quadratic approximation of the BFKL kernel
around its minimum $M=M_{\rm min}(\as)$, which in turn corresponds to the rightmost singularity
of the DL anomalous dimension.
This implies that $\gamma^{B}(N,\as)$ depends on
the intercept $c(\as)$ and curvature $\kappa(\as)$ of the kernel in
$M=M_{\rm min}$ and their derivatives with respect to $\as$.  These
are referred to as \emph{Bateman parameters}.  The Bateman parameters
are most conveniently computed using the BFKL kernel in
symmetric variables, which is related to the kernel in
DIS variables by Eq.~\eqref{eq:chisymmetry}.
Notice that  the values of $c$ and $\kappa$ are the same for both kernels.

The actual BFKL kernel used for the calculation of the Bateman parameters differs
from the one discussed in the previous subsection in two respects.  As
we explained in App.~\ref{app:double-lead-contr}, in order to cure the
unphysical branch cut arising from the eigenvalue equation of the
anomalous-dimension matrix, we did not include the full $n_f$
dependence in $\gamma_+$ (although we argued that the final result is
still correct at NLL in both $\ln (1/x)$ and $\ln Q^2$).  Here, instead,
we include the full $n_f$ dependence, since the branch-cut problem does not affect the parameters $c$ and $\kappa$.
Since the running coupling effects, though formally subleading,
are in fact dominant and determine the small-$x$ asymptotic behaviour, the inclusion
of the full $\ln Q^2$ dependence at LO and NLO is important to make sure that we do not miss relevant effects.
Note that this procedure differs
from the approach of Ref.~\cite{Altarelli:2008aj} where the rational
approximation of the $n_f$-dependent part of the eigenvalue is used
for both the DL contributions and the Bateman parameters.

The second difference between the Bateman kernel and the DL one is due to the
method used in solving the differential evolution equation.  As
observed in Sect.~\ref{sec:resgamma+}, in $M$-space the strong
coupling is a differential operator $\hat{\alpha}_s$, more precisely a
function of $-\partial/\partial M$.  In order to write the evolution
equation as a standard differential equation, the powers of
$\hat{\alpha}_s$ are moved to the left of each term.  This reordering
of operators generates further contributions due to commutators
between $\hat{\alpha}_s$ and (functions of) $M$, as described in
detail in Refs.~\cite{Ball:2005mj,Altarelli:2005ni}.

Because operator reordering of LO terms only generates NLO
contributions, at LO the Bateman (off-shell) kernel can be identified with the
fixed-coupling one
\begin{equation}
  \chi_{B,\sigma}^{\rm LO}(M,N,\as) = \chi_\sigma^{\text{LO}}(M,N,\as),
\end{equation}
constructed according to the method described in
App.~\ref{app:double-lead-contr}, except for the fact that, as we
explained, the full eigenvalue $\gamma_+$ is used in the DGLAP
contributions in $\chi_\sigma^\text{LO}$.  Because of the symmetry
$M\leftrightarrow 1-M$, this implies that, at LO, the minimum is at
$M_{\rm min}=1/2$.

At NLO, we need to include in the Bateman kernel contributions from
operator reordering of LO terms.  The position of the minimum is thus
shifted from $M=1/2$ to $M=M_{\rm min}(\as)$. The NLO Bateman off-shell kernel can
be written as
\begin{align}
  \label{eq:chiBNLO}
 \chi_{B,\sigma}^{\rm NLO}(M,N,\as) &{}= \chi_\sigma^{\text{NLO}}(M,N,\as)+\as^2\chi_1^{\beta_0}(M,N) \nonumber\\&+\chi_s^{\beta_0}(M,N,\as)+\chi_i^{\beta_0}(M,N,\as)
\end{align}
where $\chi_\sigma^{\text{NLO}}(M,N,\as)=\chi_\Sigma^{\text{NLO}}(M+N/2,N,\as)$ is the symmetric counterpart of the DL
kernel, Eq.~\eqref{eq:chiNLO} (but, once again, using the full eigenvalue of $\gamma_+^{\rm NLO}$),
$\chi_1^{\beta_0}(M,N)$ is defined in Eq.~\eqref{eq:chi1b0} and
\begin{align}
  \chi_s^{\beta_0}(M,N,\as) &= \frac{1}{2}\beta_0\frac{\as^3}{\left(M+\frac{N}{2}\right)^3}\chi_s''\left(\frac{\as}{M+\frac{N}{2}}\right) \nn
 -&\frac{1}{2}\beta_0\frac{\as^3}{\left(1-M+\frac{N}{2}\right)^3}\chi_s''\left(\frac{\as}{1-M+\frac{N}{2}}\right) \nn
-&\beta_0\frac{\as^2}{\left(1-M+\frac{N}{2}\right)^2}\chi_s'\left(\frac{\as}{1-M+\frac{N}{2}}\right), \label{eq:chisb0} \\
  \chi_i^{\beta_0}(M,N,\as) &= \frac{1}{M+N/2}\, \alpha^2\, \beta_0\,   \Bigg[ \tilde \chi_0(M,N) \nn & + \frac{1}{M+N/2}\, \chi'_s \left(\frac{\as}{M+N/2}\right) \nn &  + \frac{1}{1-M+N/2}\, \chi'_s \left(\frac{\as}{1-M+N/2}\right) \Bigg]. \label{eq:chi1beta0} 
\end{align}
The expressions in Eq.~\eqref{eq:chisb0} and~\eqref{eq:chi1beta0} are
new, since in Ref.~\cite{Altarelli:2005ni} they were given as Taylor
expansions around $M=1/2$ (notice that
$M_{\rm min}-1/2=\mathcal{O}(\as)$).  It is worth pointing out that
these expressions, on top of removing the truncation error present in
the mentioned Taylor expansion, can be easily evaluated numerically
since multiple derivatives of $\chi_s$ can always be written in terms
of derivatives of $\gamma_+^{\rm LO}$ evaluated at the solution for $\chi_s$.

The result for the Bateman anomalous dimension can be written as
\begin{align} \label{eq:gammaBU}
&  \gamma^{B}(N,\as)= \frac{1}{2}-\beta_{0}\bar{\alpha}_s  +\frac{1}{A(N,\as)} \times \nn &\times \left(\frac{2 B(N,\as)\, U(1-B(N,\as),1,z)}{U(-B(N,\as),0,z)}-1\right)\Bigg|_{z=\frac{2}{\beta_0\bar \alpha_s A(N,\as)}},
\end{align}
where $U(a,b,z)$ is the confluent hypergeometric function of the
second kind and
\begin{align} A(N,\as)\equiv{}&\sqrt{\frac{\frac{1}{2}\bar{\kappa}(\as)}{N-\bar{c}(\as)}} \\ B(N,\as)\equiv{}&\frac{1}{2\beta_{0}}\left(\frac{c'(\as)}{N-\bar{c}(\as)}+\frac{\kappa'(\as)}{\bar{\kappa}(\as)}\right)\sqrt{\frac{N-\bar{c}(\as)}{\frac{1}{2}\bar{\kappa}(\as)}} \\
  \bar c(\as)\equiv{}& c(\as)-\as c'(\as) \\
  \bar \kappa(\as)\equiv{} & \kappa(\as)-\as\kappa'(\as) \\ \frac{1}{\bar\alpha_s}\equiv{}&\frac{1}{\as}+\frac{\kappa'(\as)}{\bar{\kappa}(\as)}.
\end{align}
Notice that the only difference between $\gamma^{B,\rm LL}$ and
$\gamma^{B,\rm NLL}$ is the kernel used for the computation of the
Bateman parameters, as explained above, while their functional form is
identical.  An equivalent representation in terms of Bateman functions
(from which the name for $\gamma^B$) is given in
Ref.~\cite{Altarelli:2005ni}.

The double-counting terms between the
Bateman anomalous dimension and the DL expansion which appear in
Eq.~\eqref{eq:gammaLOrc} and~\eqref{eq:gammaNLOrc} can be written as
\begin{align} \label{eq:gammaLOrcdc}
\gamma^\text{LO,LL d.c.}(N,\as)
  &= \gamma^{B,\rm LL}_s(N,\as)+\gamma^{B,\rm LL}_{ss,0}(N,\as) \nn
  & +\gamma_{\text{match}}^{\rm LO+LL}(N,\as)+\gamma_{\text{mom}}^{\rm LO+LL} (N,\as), \\
  \gamma^\text{NLO,NLL d.c.}(N,\as)  \label{eq:gammaNLOrcdc}
  &= \gamma^{B,\rm NLL}_s(N,\as)+\gamma^{B,\rm NLL}_{ss,0}(N,\as) \nn &+\gamma^B_{ss,1}(N,\as) +\gamma_{\text{match}}^{\rm NLO+NLL} (N,\as)\nn
  & +\gamma_{\text{mom}}^{\rm NLO+NLL} (N,\as),
\end{align}
at LO and NLO respectively, where
 \begin{align} \gamma^B_s(N,\as)={}&\frac{1}{2}-\sqrt{\frac{N-c(\as)}{\frac{1}{2}\kappa(\as)}}, \\
\gamma^B_{ss,0}(N,\as)={}&-\beta_{0}\as+\frac{3}{4}\as^{2}\beta_{0}\frac{\kappa'(\as)}{\kappa(\as)}, \\ \gamma^B_{ss,1}(N,\as)={}&\frac{1}{4}\as^2\beta_0\frac{c'(\as)}{c(\as)-N} .
\end{align}
The $\gamma_{\text{match}}$ term in Eq.~\eqref{eq:gammaLOrcdc}
and~\eqref{eq:gammaNLOrcdc} removes a subleading spurious branch-cut
due to using different kernels for the DL and the Bateman anomalous dimensions.
It can be chosen to be of the form
\begin{align}
&  \gamma_{\text{match}}(N,\as) =\sqrt{\frac{N-c}{\frac{1}{2}\kappa}}-\sqrt{\frac{N-c^{\beta_0}}{\frac{1}{2}\kappa^{\beta_0}}}  -\sqrt{\frac{N+1}{\frac{1}{2}\kappa}}\nn
  &\quad +\sqrt{\frac{N+1}{\frac{1}{2}\kappa^{\beta_0}}}+\frac{1+c}{\sqrt{2\kappa(N+1)}}-\frac{1+c^{\beta_0}}{\sqrt{2\kappa^{\beta_0}(N+1)}}, \label{eq:gammamatch}
\end{align}
where $c$ and $\kappa$ are the Bateman parameters while $c^{\beta_0}$
and $\kappa^{\beta_0}$ are the intercept and curvature in the minimum
of the final off-shell kernels defined in Eq.~\eqref{eq:chiLO} and
\eqref{eq:chifinal} for the symmetrized DL result.  Finally, the term
$\gamma_{\text{mom}}$ is a subleading contribution which enforces
momentum conservation
\begin{align}
  \gamma^{\text{(N)LO}+\text{(N)LL}}_+(1,\as) = 0,
\end{align}
and can be chosen to be of the same form as $\chi_{\text{mom}}$ in
Eq.~\eqref{eq:chimom}.

We finally observe that, for the anomalous dimension
$\gamma_+^{\text{LO+LL}'}$ defined in Eq.~\eqref{eq:gammaLOp}, the
double counting term $\gamma^\text{LO,NLL d.c.}$ has the same form of
$\gamma^\text{LO, d.c.}$ but with the Bateman parameters $c$ and
$\kappa$ computed from the NLO Bateman kernel~\eqref{eq:chiBNLO} in
order to match the singularities of $\gamma^{B,\rm NLL}$.

\section{Details on the resummation of $\gamma_{qg}$}\label{app:details_gamma_QG}

In this section we provide some detail on the resummation of $\gamma_{qg}$.
Note that what follows also applies to the resummation of $U_{qg}$, Eq.~\eqref{eq:Uqg},
and of the partonic coefficient functions.

\subsection{Borel-Pad\'e method}\label{app:Borel}

Our starting point is either Eq.~\eqref{eq:gamma_qg_h} or Eq.~\eqref{eq:gamma_qg_ss_rc},
both of which provide the resummation of $\gamma_{qg}$ in terms of the function $h(M)$
which is not known in closed form.
Only the first few coefficients of its Taylor expansion in $M$, Eq.~\eqref{eq:hqgser}, are known.
However, the usage of a truncated series will inevitably decrease the all-order logarithmic accuracy
to a fixed-order accuracy. Therefore, we need a method to keep the all-order nature of the result,
while dealing with just a finite set of coefficients.

The idea used here (originally proposed in Ref.~\cite{Bonvini:2012sh}) is to construct a Pad\'e
approximant of the sum of the series from a given number of coefficients.
In practice, given that the series is expected to diverge~\cite{Altarelli:2008aj},
the actual implementation consists in summing the series \`a la Borel,
using a Pad\'e approximant for the Borel-transformed series.
Namely, Eq.~\eqref{eq:gamma_qg_ss_rc} becomes
\beq\label{eq:BorelPade1}
\gamma^{\rm NLL}_{qg} = \as \int_0^\infty dw \, e^{-w} \sum_{k=0}^\infty h_k\[\(\gamma_+^{\rm LL'}\)^k\] \frac{w^k}{k!}
\eeq
where the inner sum is to be replaced by its Pad\'e approximant.
In Ref.~\cite{Bonvini:2012sh} a diagonal $[p/p]$ Pad\'e, in which the degree of the numerator is identical
to the degree of the denominator, was used. Here, we have found that a better numerical stability is achieved
by using an almost-diagonal approximant $[p/(p-1)]$, where the degree of the denominator is lower by a unity.
We also consider a stronger second-order Borel summation, which basically consists in applying the Borel
method twice, leading to~\cite{Altarelli:2008aj,Bonvini:2012sh}
\beq\label{eq:BorelPade2}
\gamma^{\rm NLL}_{qg} = \as \int_0^\infty dw \, 2K_0(2\sqrt{w}) \sum_{k=0}^\infty h_k\[\(\gamma_+^{\rm LL'}\)^k\] \frac{w^k}{(k!)^2},
\eeq
where $K_0$ is a modified Bessel function of second kind.
Again, the inner sum is to be replaced by its Pad\'e approximant.
While this stronger method might not be strictly necessary, we find that it performs better than the first-order method.
Therefore, we use Eq.~\eqref{eq:BorelPade2} for all applications presented in this work.

This Borel-Pad\'e method, though far from optimal, works reasonably well,
provided $\as$ is not too large ($\as\lesssim 0.3$)
and the number of coefficient used is not too high (we use $p=8$, i.e.\ 16 coefficients).
We adopt this method also for the computation of the part of $U_{qg}$, Eq.~\eqref{eq:Uqg},
which is given as a series.

\subsection{Large-$N$ subtraction}
\label{app:gammaqgcomparison}

We now briefly comment on the two ways of subtracting the large-$N$ behaviour in the computation
of $\gamma_{qg}$ we mentioned in Sect.~\ref{sec:resgammaqg}.
For ease of notation, let us denote
\beq
\gamma = \gamma_+^{\rm LO+LL'},\qquad
\bar\gamma = \gamma_+^{\rm LO} - \gamma_+^{\rm LO,sing},
\eeq
such that $\gamma_+^{\rm LL'}=\gamma-\bar\gamma$, and
\beq\label{eq:gammagamma0limit}
\lim_{N\to\infty}(\gamma-\bar\gamma) = 0.
\eeq
Ignoring for simplicity the complications coming from the resummation of running coupling effects,
which is not relevant for the present discussion, we implement the resummation of $\gamma_{qg}$ as
\beq
\gamma_{qg} = \as h\(\gamma-\bar\gamma\),
\eeq
Eq.~\eqref{eq:gamma_qg_ss_rc}, which automatically vanishes in the large-$N$ limit due to Eq.~\eqref{eq:gammagamma0limit}
(except for the 0-th order term of the series, which is anyway subtracted when matching to fixed order).
This differs from the choice performed by ABF~\cite{Altarelli:2008aj}\footnote
{We observe that Eq.~(3.29) and (4.25) of Ref.~\cite{Altarelli:2008aj} contain several typos.}
\beq\label{eq:richard}
\gamma_{qg}^{\rm ABF} = \as h(\gamma) - \as h(\bar\gamma) + \as h(0),
\eeq
where the first term contains the small-$x$ resummation,
and the second term subtracts the large-$N$ behaviour by recomputing $h$ with $\bar\gamma$ as argument;
finally, the zero-th order constant is restored with the last term.\footnote
{The last term is actually irrelevant when computing just $\Delta\gamma_{qg}$, as it is subtracted out anyway.}
Note that in the original ABF formulation the full NLO+NLL anomalous dimension was used,
however here we are interested in the different ways of subtracting the large-$N$ behaviour,
so we do not need to add this complication to the discussion.
In other words, we apply the large-$N$ subtraction \emph{before} acting with $h$, while ABF do it \emph{after} $h$.
Our option is closer to the ``plain'' resummation obtained by $\gamma_{qg}^s = h(\gamma_s)$,
and has the advantage of having to compute $h$ (through the Borel-Pad\'e method) only once.

To understand the differences between the two approaches we expand the two results
\begin{align}
  \gamma_{qg} &= 1+ h_1(\gamma-\bar\gamma)  + h_2(\gamma^2+\bar\gamma^2-2\gamma\bar\gamma)
                \nn
  &\quad + h_3(\gamma^3-\bar\gamma^3-3\gamma^2\bar\gamma+3\gamma\bar\gamma^2) +\ldots\\
  \gamma_{qg}^{\rm ABF} &= 1+ h_1(\gamma-\bar\gamma)  + h_2(\gamma^2-\bar\gamma^2)  \nn
  &\quad + h_3(\gamma^3-\bar\gamma^3) +\ldots
\end{align}
from which we can write the difference as
\begin{align}\label{eq:difference}
&\gamma_{qg}^{\rm ABF} -\gamma_{qg}= \bar\gamma(\gamma-\bar\gamma)
\Bigg[ 2h_2 + 3h_3 \gamma + \ldots
\nn
  &\quad +h_k \sum_{j=0}^{k-2}\bar\gamma^j\gamma^{k-2-j}\[1+(-1)^j\binom{k-1}{j+1}\]
+\ldots \Bigg]
\end{align}
These terms vanish at large $N$ because of Eq.~\eqref{eq:gammagamma0limit}, as they should,
so the large-$x$ limit is the same with the two procedures.
At small $N$, close to the pole, these terms vanish only if $\bar\gamma$ vanishes in $N=0$.
When this is the case, it is then clear that the two approaches will give equivalent results
(this is indeed what we find). However, if $\bar\gamma$ does not vanish in $N=0$,
the difference, though formally subleading, can be sizeable.

From this observation it seems favourable to construct $\bar\gamma$ such that it vanishes
in $N=0$. This is achieved if $\gamma_+^{\rm LO,sing}$ contains not only LL terms (as formally strictly necessary)
but also NLL terms (which are not formally needed for the present accuracy).
Expanding the LO at small $N$ up to NLL we have
\beq\label{eq:LOsing}
\gamma_+^{\rm LO,sing} = \frac{\as}{\pi}\left(  \frac{C_A}{N} - \frac{11 C_A+2n_f(1-2C_F/C_A)}{12 (N+1)}\right).
\eeq
Note that in the second term (the NLL contribution) we have added a damping factor $1/(N+1)$.
This is needed because this NLL term is originally a constant, and therefore if included without damping it would spoil
Eq.~\eqref{eq:gammagamma0limit}, and hence the large-$N$ subtraction.

\section{Details of DIS resummation}\label{app:dis}

In this appendix we collect the relevant expressions for the massless off-shell coefficient functions in DIS,
needed for the resummation described in Sect.~\ref{sec:coeff-func}, and discuss the matching to fixed order.

\subsection{Massless off-shell coefficient functions}

The off-shell cross section in the case of massive quarks has been computed in Ref~\cite{Catani:1990eg};
more precisely, the $N=0$ moment of the cross section is Eq~(4.12).\footnote
{$N$ moments are computed with respect to $\rho=\frac{4m^2}{s}$ rather than $z=\frac{Q^2}{s}$, however the difference is subleading.}
Here we take the massless limit of the expression reported in Ref.~\cite{Catani:1990eg}, obtaining
\begin{align} \label{eq:C2off}
  &{\cal C}_2(0,\xi,\as)
  = n_f\frac{\as}{3\pi}\, \frac38 \int_0^1\frac{dx_1}{\sqrt{1-x_1}} \int_0^1\frac{dx_2}{\sqrt{1-x_2}}\, \nn
  & \frac{(2-x_1) x_2^2+ x_1 x_2 \xi (3 x_1+3 x_2-4 x_1 x_2)+(2-x_2)x_1^2 \xi^2}{(x_2+x_1\xi)^3}.
\end{align}
Its Mellin transform is~\cite{Catani:1994sq}
\begin{align} \label{eq:C2offMellin}
  &\tilde{\cal C}_2(0,M,\as)
  = M\int_0^\infty d\xi\, \xi^{M-1} {\cal C}_2(0,\xi,\as) \nonumber \\
  &=  n_f\frac{\as}{3\pi}\, \frac1M\, \frac{3(2+3 M-3 M^2)}{2(3-2M)} \frac{\Gamma^3(1-M)\Gamma^3(1+M)}{\Gamma(2-2M)\Gamma(2+2M)}.
\end{align}
For the longitudinal coefficient function, we find an expression similar to Eq.~\eqref{eq:C2off}
\begin{align} \label{eq:CLoff}
  {\cal C}_L(0,\xi,\as)
&  = n_f \frac{\as}{3\pi}\, \frac38 \int_0^1\frac{dx_1}{\sqrt{1-x_1}} \int_0^1\frac{dx_2}{\sqrt{1-x_2}}\, \nn
  &\times\frac{x_1x_2}{x_1+x_2\xi};
\end{align}
its Mellin transform reproduces the result of Ref.~\cite{Catani:1994sq}:
\begin{align} \label{eq:CLoffMellin}
  &\tilde {\cal C}_L(0,M,\as)
  = M\int_0^\infty d\xi\, \xi^{M-1} {\cal C}_L(0,\xi,\as) \nonumber\\
  &= n_f\frac{\as}{3\pi}\, \frac{3(1-M)}{3-2M} \frac{\Gamma^3(1-M)\Gamma^3(1+M)}{\Gamma(2-2M)\Gamma(2+2M)}. 
\end{align}
For our numerical implementation, we find useful to note that both ${\cal C}_2(0,\xi,\as)$ and ${\cal C}_L(0,\xi,\as)$ satisfy
\beq
{\cal C}(0,\xi,\as) = \frac1\xi\, {\cal C}\(0,\frac1\xi,\as\),
\eeq
which implies
\beq
\frac{\tilde{\cal C}(0,M,\as)}{M} = \frac{\tilde{\cal C}(0,1-M,\as)}{1-M}.
\eeq

\subsection{Matching to fixed-order}
\label{app:disexpansion}
In order to calculate the resummation contributions $\Delta_nC_{i}$, $i=g,q$,
defined in Eq.~\eqref{eq:deltaC}, we have to consider the perturbative expansion of the resummed results.
Regardless of the approximation we use to compute it, the evolution factor $U$ can be expanded as
\beq\label{eq:Uexpanded}
U(N,\xi) = 1+\as\gamma_+^{(0)}(N)\ln\xi +\Ord(\as^2).
\eeq
In our case the resummed anomalous dimension is the LL$^\prime$ one, Eq.~\eqref{eq:gammaLLp}, introduced in Sect.~\ref{sec:resgammaqg},
so $\as\gamma_+^{(0)}=\gamma_+^{\rm LO,sing}$, Eq.~\eqref{eq:LOsing}.

In the case of the longitudinal structure function $F_L$, we can plug Eq.~\eqref{eq:Uexpanded}
into Eq.~\eqref{eq:res_CLfinal} and get
\begin{align}
C_{L,g}(N,\as)
  &= \as C_{L,g}^{(1)}(N) + \as^2 C_{L,g}^{(2)}(N) + \Ord(\as^3),
\end{align}
with
\begin{align}\label{eq:CLexpansion}
  C_{L,g}^{(1)}(N) &= {\cal C}_L\(0,0,1\) 
                     , \nonumber\\
  C_{L,g}^{(2)}(N) &= - \gamma_+^{(0)}(N) \int_{0}^\infty d\xi \, \frac{d}{d\xi}{\cal C}_L\(0,\xi,1\)\, \ln\xi,
\end{align}
where we made explicit use of the fact that ${\cal C}\(N,\xi,\as\)$ is linear in $\as$,
and we let $\xi_0\to0$ since we are expanding to fixed order.
The expansion of $C_2$,
\begin{align}
C_{2,g}(N,\as)
  &= \as C_{2,g}^{(1)}(N) + \as^2 C_{2,g}^{(2)}(N) + \Ord(\as^3),
\end{align}
is obtained equivalently by plugging Eq.~\eqref{eq:Uexpanded} into Eq.~\eqref{eq:C2resfinal}.
The solution is more involved and reads
\begin{align}
C_{2,g}^{(1)}(N) &= \Big({\cal C}_2\(0,\xi,1\)+h_0\ln\xi\Big)_{\xi=0} - h_1 , \nonumber\\
C_{2,g}^{(2)}(N) &= - \gamma_+^{(0)}(N) \Bigg[\int_{0}^\infty d\xi \, \Big(\frac{d}{d\xi}{\cal C}_2\(0,\xi,1\)\nn
                 &\qquad\qquad\qquad +\frac{h_0}\xi\theta(1-\xi) \Big) \ln\xi +h_2\Bigg].
\end{align}

Having the expansion of the resummed coefficient functions up to $\Ord(\as^2)$, we can
now construct both $\Delta_1C_{i}$ and $\Delta_2C_{i}$ in $N$ space, and then in $x$ space by Mellin inversion.
We have however noted that, while $\Delta_2C_{i}(x,\as)$, which contains contributions starting at $\Ord(\as^3)$,
vanishes fast enough at large $x$ (after applying the $(1-x)^2$ damping discussed in Sect.~\ref{sec:cfnumeric}) and hence ensures a smooth matching onto the fixed order,
$\Delta_1C_{i}(x,\as)$, which contains contributions starting at $\Ord(\as^2)$, is sizeable at large $x\sim10^{-1}$,
thereby potentially spoiling the accuracy of the resummed and matched NLO+LL result in that region.
Since the culprit of this sizable effect is exactly the $\Ord(\as^2)$ term of the expanded resummation,
we find it convenient to re-define $\Delta_1C_{i}(x,\as)$ as
\begin{align}\label{eq:alterD1}
\Delta_1 C_{i}(x,\as) &\equiv \Delta_2 C_{i}(x,\as)+ \as^2 C_{i}^{(2)}(x) (1-x)^2 f(x),
\end{align}
where the damping $(1-x)^2$ is the standard damping adopted everywhere,
and $f(x)$ is a further damping function such that $f(0)=1$ and $f(1)=0$.
We have identified a convenient form for the damping function in $f(x) = (1-\sqrt{x})^6$.
We observe that for $f(x) = 1$ one would recover the original definition.

\phantomsection
\addcontentsline{toc}{section}{References}

\bibliographystyle{jhep}
\bibliography{references}

\providecommand{\href}[2]{#2}\begingroup\raggedright\begin{thebibliography}{10}

\bibitem{Abramowicz:2015mha}
{\bf ZEUS, H1} Collaboration, H.~Abramowicz et~al., {\it {Combination of
  measurements of inclusive deep inelastic ${e^{\pm }p}$ scattering cross
  sections and QCD analysis of HERA data}},  {\em Eur. Phys. J.} {\bf C75}
  (2015), no.~12 580, [\href{http://arxiv.org/abs/1506.06042}{{\tt
  arXiv:1506.06042}}].

\bibitem{Caola:2009iy}
F.~Caola, S.~Forte, and J.~Rojo, {\it {Deviations from NLO QCD evolution in
  inclusive HERA data}},  {\em Phys. Lett.} {\bf B686} (2010) 127--135,
  [\href{http://arxiv.org/abs/0910.3143}{{\tt arXiv:0910.3143}}].

\bibitem{Caola:2010cy}
F.~Caola, S.~Forte, and J.~Rojo, {\it {HERA data and DGLAP evolution: Theory
  and phenomenology}},  {\em Nucl. Phys.} {\bf A854} (2011) 32--44,
  [\href{http://arxiv.org/abs/1007.5405}{{\tt arXiv:1007.5405}}].

\bibitem{Gribov:1972ri}
V.~N. Gribov and L.~N. Lipatov, {\it {Deep inelastic e p scattering in
  perturbation theory}},  {\em Sov. J. Nucl. Phys.} {\bf 15} (1972) 438--450.
  [Yad. Fiz.15,781(1972)].

\bibitem{Dokshitzer:1977sg}
Y.~L. Dokshitzer, {\it {Calculation of the Structure Functions for Deep
  Inelastic Scattering and e+ e- Annihilation by Perturbation Theory in Quantum
  Chromodynamics.}},  {\em Sov. Phys. JETP} {\bf 46} (1977) 641--653. [Zh.
  Eksp. Teor. Fiz.73,1216(1977)].

\bibitem{Altarelli:1977zs}
G.~Altarelli and G.~Parisi, {\it {Asymptotic Freedom in Parton Language}},
  {\em Nucl. Phys.} {\bf B126} (1977) 298--318.

\bibitem{Lipatov:1976zz}
L.~N. Lipatov, {\it {Reggeization of the Vector Meson and the Vacuum
  Singularity in Nonabelian Gauge Theories}},  {\em Sov. J. Nucl. Phys.} {\bf
  23} (1976) 338--345.

\bibitem{Fadin:1975cb}
V.~S. Fadin, E.~Kuraev, and L.~Lipatov, {\it {On the Pomeranchuk Singularity in
  Asymptotically Free Theories}},  {\em Phys.Lett.} {\bf B60} (1975) 50--52.

\bibitem{Kuraev:1976ge}
E.~A. Kuraev, L.~N. Lipatov, and V.~S. Fadin, {\it {Multi - Reggeon Processes
  in the Yang-Mills Theory}},  {\em Sov.Phys.JETP} {\bf 44} (1976) 443--450.

\bibitem{Kuraev:1977fs}
E.~A. Kuraev, L.~N. Lipatov, and V.~S. Fadin, {\it {The Po\-me\-ran\-chuk
  Singularity in Nonabelian Gauge Theories}},  {\em Sov. Phys. JETP} {\bf 45}
  (1977) 199--204.

\bibitem{Balitsky:1978ic}
I.~I. Balitsky and L.~N. Lipatov, {\it {The Pomeranchuk Singularity in Quantum
  Chromodynamics}},  {\em Sov. J. Nucl. Phys.} {\bf 28} (1978) 822--829.

\bibitem{Fadin:1998py}
V.~S. Fadin and L.~Lipatov, {\it {BFKL pomeron in the next-to-leading
  approximation}},  {\em Phys.Lett.} {\bf B429} (1998) 127--134,
  [\href{http://arxiv.org/abs/hep-ph/9802290}{{\tt hep-ph/9802290}}].

\bibitem{Salam:1998tj}
G.~Salam, {\it {A Resummation of large subleading corrections at small x}},
  {\em JHEP} {\bf 9807} (1998) 019,
  [\href{http://arxiv.org/abs/hep-ph/9806482}{{\tt hep-ph/9806482}}].

\bibitem{Ciafaloni:1999yw}
M.~Ciafaloni, D.~Colferai, and G.~Salam, {\it {Renormalization group improved
  small x equation}},  {\em Phys.Rev.} {\bf D60} (1999) 114036,
  [\href{http://arxiv.org/abs/hep-ph/9905566}{{\tt hep-ph/9905566}}].

\bibitem{Ciafaloni:2003rd}
M.~Ciafaloni, D.~Colferai, G.~Salam, and A.~Stasto, {\it {Renormalization group
  improved small x Green's function}},  {\em Phys.Rev.} {\bf D68} (2003)
  114003, [\href{http://arxiv.org/abs/hep-ph/0307188}{{\tt hep-ph/0307188}}].

\bibitem{Ciafaloni:2007gf}
M.~Ciafaloni, D.~Colferai, G.~Salam, and A.~Stasto, {\it {A Matrix formulation
  for small-$x$ singlet evolution}},  {\em JHEP} {\bf 0708} (2007) 046,
  [\href{http://arxiv.org/abs/0707.1453}{{\tt arXiv:0707.1453}}].

\bibitem{Ball:1995vc}
R.~D. Ball and S.~Forte, {\it {Summation of leading logarithms at small x}},
  {\em Phys.Lett.} {\bf B351} (1995) 313--324,
  [\href{http://arxiv.org/abs/hep-ph/9501231}{{\tt hep-ph/9501231}}].

\bibitem{Ball:1997vf}
R.~D. Ball and S.~Forte, {\it {Asymptotically free partons at high-energy}},
  {\em Phys.Lett.} {\bf B405} (1997) 317--326,
  [\href{http://arxiv.org/abs/hep-ph/9703417}{{\tt hep-ph/9703417}}].

\bibitem{Altarelli:2001ji}
G.~Altarelli, R.~D. Ball, and S.~Forte, {\it {Factorization and resummation of
  small x scaling violations with running coupling}},  {\em Nucl.Phys.} {\bf
  B621} (2002) 359--387, [\href{http://arxiv.org/abs/hep-ph/0109178}{{\tt
  hep-ph/0109178}}].

\bibitem{Altarelli:2003hk}
G.~Altarelli, R.~D. Ball, and S.~Forte, {\it {An Anomalous dimension for small
  x evolution}},  {\em Nucl.Phys.} {\bf B674} (2003) 459--483,
  [\href{http://arxiv.org/abs/hep-ph/0306156}{{\tt hep-ph/0306156}}].

\bibitem{Altarelli:2005ni}
G.~Altarelli, R.~D. Ball, and S.~Forte, {\it {Perturbatively stable resummed
  small x evolution kernels}},  {\em Nucl.Phys.} {\bf B742} (2006) 1--40,
  [\href{http://arxiv.org/abs/hep-ph/0512237}{{\tt hep-ph/0512237}}].

\bibitem{Altarelli:2008aj}
G.~Altarelli, R.~D. Ball, and S.~Forte, {\it {Small x Resummation with Quarks:
  Deep-Inelastic Scattering}},  {\em Nucl.Phys.} {\bf B799} (2008) 199--240,
  [\href{http://arxiv.org/abs/0802.0032}{{\tt arXiv:0802.0032}}].

\bibitem{Rojo:2009us}
J.~Rojo, G.~Altarelli, R.~D. Ball, and S.~Forte, {\it {Towards small x resummed
  DIS phenomenology}},  in {\em {Proceedings, 17th International Workshop on
  Deep-Inelastic Scattering and Related Subjects (DIS 2009)}}, 2009.
\newblock \href{http://arxiv.org/abs/0907.0443}{{\tt arXiv:0907.0443}}.

\bibitem{Thorne:1999sg}
R.~S. Thorne, {\it {Explicit calculation of the running coupling BFKL anomalous
  dimension}},  {\em Phys. Lett.} {\bf B474} (2000) 372--384,
  [\href{http://arxiv.org/abs/hep-ph/9912284}{{\tt hep-ph/9912284}}].

\bibitem{Thorne:1999rb}
R.~S. Thorne, {\it {NLO BFKL equation, running coupling and renormalization
  scales}},  {\em Phys. Rev.} {\bf D60} (1999) 054031,
  [\href{http://arxiv.org/abs/hep-ph/9901331}{{\tt hep-ph/9901331}}].

\bibitem{Thorne:2001nr}
R.~S. Thorne, {\it {The Running coupling BFKL anomalous dimensions and
  splitting functions}},  {\em Phys. Rev.} {\bf D64} (2001) 074005,
  [\href{http://arxiv.org/abs/hep-ph/0103210}{{\tt hep-ph/0103210}}].

\bibitem{White:2006yh}
C.~D. White and R.~S. Thorne, {\it {A Global Fit to Scattering Data with NLL
  BFKL Resummations}},  {\em Phys. Rev.} {\bf D75} (2007) 034005,
  [\href{http://arxiv.org/abs/hep-ph/0611204}{{\tt hep-ph/0611204}}].

\bibitem{Rothstein:2016bsq}
I.~Z. Rothstein and I.~W. Stewart, {\it {An Effective Field Theory for Forward
  Scattering and Factorization Violation}},
  \href{http://arxiv.org/abs/1601.04695}{{\tt arXiv:1601.04695}}.

\bibitem{Catani:1990xk}
S.~Catani, M.~Ciafaloni, and F.~Hautmann, {\it {Gluon contributions to
  small-$x$ heavy flavor production}},  {\em Phys.Lett.} {\bf B242} (1990) 97.

\bibitem{Catani:1990eg}
S.~Catani, M.~Ciafaloni, and F.~Hautmann, {\it {High energy factorization and
  small-$x$ heavy flavour production}},  {\em Nucl. Phys.} {\bf B366} (1991)
  135--188.

\bibitem{Collins:1991ty}
J.~C. Collins and R.~K. Ellis, {\it {Heavy quark production in very high energy
  hadron collisions}},  {\em Nucl. Phys.} {\bf B360} (1991) 3--30.

\bibitem{Catani:1993ww}
S.~Catani, M.~Ciafaloni, and F.~Hautmann, {\it {High-energy factorization in
  QCD and minimal subtraction scheme}},  {\em Phys.Lett.} {\bf B307} (1993)
  147--153.

\bibitem{Catani:1993rn}
S.~Catani and F.~Hautmann, {\it {Quark anomalous dimensions at small x}},  {\em
  Phys.Lett.} {\bf B315} (1993) 157--163.

\bibitem{Catani:1994sq}
S.~Catani and F.~Hautmann, {\it {High-energy factorization and small x deep
  inelastic scattering beyond leading order}},  {\em Nucl.Phys.} {\bf B427}
  (1994) 475--524, [\href{http://arxiv.org/abs/hep-ph/9405388}{{\tt
  hep-ph/9405388}}].

\bibitem{Ball:2007ra}
R.~D. Ball, {\it {Resummation of Hadroproduction Cross-sections at High
  Energy}},  {\em Nucl.Phys.} {\bf B796} (2008) 137--183,
  [\href{http://arxiv.org/abs/0708.1277}{{\tt arXiv:0708.1277}}].

\bibitem{Caola:2010kv}
F.~Caola, S.~Forte, and S.~Marzani, {\it {Small x resummation of rapidity
  distributions: The Case of Higgs production}},  {\em Nucl.Phys.} {\bf B846}
  (2011) 167--211, [\href{http://arxiv.org/abs/1010.2743}{{\tt
  arXiv:1010.2743}}].

\bibitem{Ball:2001pq}
R.~Ball and R.~K. Ellis, {\it {Heavy quark production at high-energy}},  {\em
  JHEP} {\bf 0105} (2001) 053, [\href{http://arxiv.org/abs/hep-ph/0101199}{{\tt
  hep-ph/0101199}}].

\bibitem{Marzani:2008uh}
S.~Marzani and R.~D. Ball, {\it {High Energy Resummation of Drell-Yan
  Processes}},  {\em Nucl.Phys.} {\bf B814} (2009) 246--264,
  [\href{http://arxiv.org/abs/0812.3602}{{\tt arXiv:0812.3602}}].

\bibitem{Diana:2009xv}
G.~Diana, {\it {High-energy resummation in direct photon production}},  {\em
  Nucl. Phys.} {\bf B824} (2010) 154--167,
  [\href{http://arxiv.org/abs/0906.4159}{{\tt arXiv:0906.4159}}].

\bibitem{Diana:2010ef}
G.~Diana, J.~Rojo, and R.~D. Ball, {\it {High energy resummation of direct
  photon production at hadronic colliders}},  {\em Phys.Lett.} {\bf B693}
  (2010) 430--437, [\href{http://arxiv.org/abs/1006.4250}{{\tt
  arXiv:1006.4250}}].

\bibitem{Hautmann:2002tu}
F.~Hautmann, {\it {Heavy top limit and double logarithmic contributions to
  Higgs production at $m_H^2/s$ much less than 1}},  {\em Phys.Lett.} {\bf
  B535} (2002) 159--162, [\href{http://arxiv.org/abs/hep-ph/0203140}{{\tt
  hep-ph/0203140}}].

\bibitem{Pasechnik:2006du}
R.~S. Pasechnik, O.~V. Teryaev, and A.~Szczurek, {\it {Scalar Higgs boson
  production in a fusion of two off-shell gluons}},  {\em Eur. Phys. J.} {\bf
  C47} (2006) 429--435, [\href{http://arxiv.org/abs/hep-ph/0603258}{{\tt
  hep-ph/0603258}}].

\bibitem{Marzani:2008az}
S.~Marzani, R.~D. Ball, V.~Del~Duca, S.~Forte, and A.~Vicini, {\it {Higgs
  production via gluon-gluon fusion with finite top mass beyond next-to-leading
  order}},  {\em Nucl.Phys.} {\bf B800} (2008) 127--145,
  [\href{http://arxiv.org/abs/0801.2544}{{\tt arXiv:0801.2544}}].

\bibitem{Caola:2011wq}
F.~Caola and S.~Marzani, {\it {Finite fermion mass effects in pseudoscalar
  Higgs production via gluon-gluon fusion}},  {\em Phys.Lett.} {\bf B698}
  (2011) 275--283, [\href{http://arxiv.org/abs/1101.3975}{{\tt
  arXiv:1101.3975}}].

\bibitem{Forte:2015gve}
S.~Forte and C.~Muselli, {\it {High energy resummation of transverse momentum
  distributions:Higgs in gluon fusion}},
  \href{http://arxiv.org/abs/1511.05561}{{\tt arXiv:1511.05561}}.

\bibitem{Ciafaloni:2005cg}
M.~Ciafaloni and D.~Colferai, {\it {Dimensional regularisation and
  factorisation schemes in the BFKL equation at subleading level}},  {\em JHEP}
  {\bf 09} (2005) 069, [\href{http://arxiv.org/abs/hep-ph/0507106}{{\tt
  hep-ph/0507106}}].

\bibitem{Marzani:2007gk}
S.~Marzani, R.~D. Ball, P.~Falgari, and S.~Forte, {\it {BFKL at
  next-to-next-to-leading order}},  {\em Nucl. Phys.} {\bf B783} (2007)
  143--175, [\href{http://arxiv.org/abs/0704.2404}{{\tt arXiv:0704.2404}}].

\bibitem{Caron-Huot:2016tzz}
S.~Caron-Huot and M.~Herranen, {\it {High-energy evolution to three loops}},
  \href{http://arxiv.org/abs/1604.07417}{{\tt arXiv:1604.07417}}.

\bibitem{Dittmar:2009ii}
M.~Dittmar et~al., {\it {Parton Distributions}},
  \href{http://arxiv.org/abs/0901.2504}{{\tt arXiv:0901.2504}}.

\bibitem{Jaroszewicz:1982gr}
T.~Jaroszewicz, {\it {Gluonic Regge Singularities and Anomalous Dimensions in
  QCD}},  {\em Phys. Lett.} {\bf B116} (1982) 291.

\bibitem{Catani:1989sg}
S.~Catani, F.~Fiorani, and G.~Marchesini, {\it {Small-$x$ behavior of initial
  state radiation in perturbative QCD}},  {\em Nucl. Phys.} {\bf B336} (1990)
  18--85.

\bibitem{Altarelli:1999vw}
G.~Altarelli, R.~D. Ball, and S.~Forte, {\it {Resummation of singlet parton
  evolution at small x}},  {\em Nucl. Phys.} {\bf B575} (2000) 313--329,
  [\href{http://arxiv.org/abs/hep-ph/9911273}{{\tt hep-ph/9911273}}].

\bibitem{Bertone:2013vaa}
V.~Bertone, S.~Carrazza, and J.~Rojo, {\it {APFEL: A PDF Evolution Library with
  QED corrections}},  {\em Comput. Phys. Commun.} {\bf 185} (2014) 1647--1668,
  [\href{http://arxiv.org/abs/1310.1394}{{\tt arXiv:1310.1394}}].

\bibitem{Harlander:2009my}
R.~V. Harlander, H.~Mantler, S.~Marzani, and K.~J. Ozeren, {\it {Higgs
  production in gluon fusion at next-to-next-to-leading order QCD for finite
  top mass}},  {\em Eur.Phys.J.} {\bf C66} (2010) 359--372,
  [\href{http://arxiv.org/abs/0912.2104}{{\tt arXiv:0912.2104}}].

\bibitem{Collins:1350496}
J.~Collins, {\em {Foundations of perturbative QCD}}.
\newblock Cambridge monographs on particle physics, nuclear physics, and
  cosmology. Cambridge Univ. Press, New York, NY, 2011.

\bibitem{Vermaseren:2005qc}
J.~A.~M. Vermaseren, A.~Vogt, and S.~Moch, {\it {The Third-order QCD
  corrections to deep-inelastic scattering by photon exchange}},  {\em Nucl.
  Phys.} {\bf B724} (2005) 3--182,
  [\href{http://arxiv.org/abs/hep-ph/0504242}{{\tt hep-ph/0504242}}].

\bibitem{Ball:2014uwa}
{\bf NNPDF} Collaboration, R.~D. Ball et~al., {\it {Parton distributions for
  the LHC Run II}},  {\em JHEP} {\bf 04} (2015) 040,
  [\href{http://arxiv.org/abs/1410.8849}{{\tt arXiv:1410.8849}}].

\bibitem{Rojo:2016kwu}
J.~Rojo, {\it {Parton Distributions at a 100 TeV Hadron Collider}},  in {\em
  {24th International Workshop on Deep-Inelastic Scattering and Related
  Subjects (DIS 2016) Hamburg, Germany, April 11-15, 2016}}, 2016.
\newblock \href{http://arxiv.org/abs/1605.08302}{{\tt arXiv:1605.08302}}.

\bibitem{Ball:2013bra}
R.~D. Ball, M.~Bonvini, S.~Forte, S.~Marzani, and G.~Ridolfi, {\it {Higgs
  production in gluon fusion beyond NNLO}},  {\em Nucl.Phys.} {\bf B874} (2013)
  746--772, [\href{http://arxiv.org/abs/1303.3590}{{\tt arXiv:1303.3590}}].

\bibitem{Marzani:2015oyb}
S.~Marzani, {\it {Combining $Q_T$ and small-$x$ resummations}},  {\em Phys.
  Rev.} {\bf D93} (2016), no.~5 054047,
  [\href{http://arxiv.org/abs/1511.06039}{{\tt arXiv:1511.06039}}].

\bibitem{Ball:2005mj}
R.~D. Ball and S.~Forte, {\it {All order running coupling BFKL evolution from
  GLAP (and vice-versa)}},  {\em Nucl. Phys.} {\bf B742} (2006) 158--175,
  [\href{http://arxiv.org/abs/hep-ph/0601049}{{\tt hep-ph/0601049}}].

\bibitem{Bonvini:2012sh}
M.~Bonvini, {\em {Resummation of soft and hard gluon radiation in perturbative
  QCD}}.
\newblock PhD thesis, Genoa U., 2012.
\newblock \href{http://arxiv.org/abs/1212.0480}{{\tt arXiv:1212.0480}}.

\end{thebibliography}\endgroup

\end{document}